\documentclass[preprint,12pt]{elsarticle}
\usepackage[utf8]{inputenc}
\usepackage{textgreek}

\biboptions{numbers,sort&compress}

\usepackage{amssymb}
\usepackage{amsmath}
\usepackage[table]{xcolor}
\usepackage{mdframed}
\usepackage{xcolor}
\definecolor{lightgray}{gray}{0.95}
\newmdenv[
  backgroundcolor=lightgray,
  linewidth=0pt,
  roundcorner=2pt,
  innertopmargin=4pt,
  innerbottommargin=4pt,
  innerleftmargin=6pt,
  innerrightmargin=6pt,
  skipabove=6pt,
  skipbelow=6pt
]{keytakeaway}

\usepackage[final]{microtype}     
\usepackage[T1]{fontenc}
\usepackage{xurl}
\usepackage{tabularray}
\usepackage{adjustbox}
\usepackage{graphicx}  
\newcommand{\rot}[1]{\rotatebox[origin=c]{90}{#1}} 
\usepackage{enumitem}
\usepackage{array} 
\usepackage{float}
\usepackage{tabularx}
\usepackage[font=small,labelfont=bf,justification=centering]{caption}
\usepackage{longtable,booktabs, threeparttable}

\colorlet{HeaderGray}{gray!5}

\newlength\sidecolwidth
\usepackage{subcaption}

\newcolumntype{Y}{>{\raggedright\arraybackslash}X}
\renewcommand{\arraystretch}{1.2}
\usepackage{tikz}
\newcommand*\emptycirc[1][1ex]{\tikz\draw (0,0) circle (#1);} 
\newcommand*\halfcirc[1][1ex]{%
  \begin{tikzpicture}
  \draw[fill] (0,0)-- (90:#1) arc (90:270:#1) -- cycle ;
  \draw (0,0) circle (#1);
  \end{tikzpicture}}
\newcommand*\fullcirc[1][1ex]{\tikz\fill (0,0) circle (#1);}

\usepackage{titlesec}
\titleformat{\paragraph}[runin]{\normalfont\normalsize\bfseries}{\theparagraph}{1em}{}
\titleclass{\subsubsubsection}{straight}[\subsubsection]
\newcounter{subsubsubsection}[subsubsection]
\renewcommand\thesubsubsubsection{\thesubsubsection.\arabic{subsubsubsection}}
\titleformat{\subsubsubsection}[runin]{\normalfont\normalsize\bfseries}{\thesubsubsubsection}{1em}{}
\titlespacing*{\subsubsubsection}{0pt}{0.5em}{0.5em}

\newcommand{\score}[1]{\ifthenelse{\equal{#1}{\textbullet}}{2}{\ifthenelse{\equal{#1}{\textperiodcentered}}{1}{0}}} 

\journal{Journal of Systems and Software}

\begin{document}

\begin{frontmatter}



\title{Test Case Specification Techniques and System Testing Tools in the Automotive Industry: A Review}

\author[1,2]{Denesa Zyberaj\corref{cor1}}
\ead{denesa.zyberaj@mercedes-benz.com}
\author[1]{Pascal Hirmer}
\ead{pascal.hirmer@mercedes-benz.com}
\author[2]{Marco Aiello}
\ead{marco.aiello@iaas.uni-stuttgart.de}
\author[3]{Stefan Wagner}
\ead{stefan.wagner@tum.de}

\affiliation[1]{organization={Mercedes-Benz AG},
            city={Sindelfingen},
            country={Germany}}

\affiliation[2]{organization={University of Stuttgart},
            city={Stuttgart},
            country={Germany}}

\affiliation[3]{organization={Technical University of Munich},
            city={Heilbronn},
            country={Germany}}

\cortext[cor1]{Corresponding author. Email: denesa.zyberaj@mercedes-benz.com}

\begin{abstract}
The automotive domain is shifting to software-centric development to meet regulation, market pressure, and feature velocity. This shift increases \mbox{embedded} systems' complexity and strains testing capacity. Despite relevant standards, a coherent system-testing methodology that spans heterogeneous, legacy-constrained toolchains remains elusive, and practice often depends on individual expertise rather than a systematic strategy. We derive challenges and requirements from a systematic literature review (SLR), complemented by industry experience and practice. We map them to test case specification techniques and testing tools, evaluating their suitability for automotive testing using PRISMA. Our contribution is a curated catalog that supports technique/tool selection and can inform future testing frameworks and improvements. We synthesize nine recurring challenge areas across the life cycle, such as requirements quality and traceability, variability management, and toolchain fragmentation. We then provide a prioritized criteria catalog that recommends model-based planning, interoperable and traceable toolchains, requirements uplift, pragmatic automation and virtualization, targeted AI and formal methods, actionable metrics, and lightweight organizational practices.
\end{abstract}



\begin{keyword}
Testing \sep Automotive \sep Testing Tools\sep Requirements \sep Specification \sep Literature Review 


\end{keyword}

\end{frontmatter}



\section{Introduction}
\label{intro}
The automotive industry is experiencing a major shift. Next to traditional areas of innovation and R\&D such as mechanical engineering, engine development, aerodynamics, materials science, and ergonomics, the industry is seeing an increasingly significant role for software and software-driven development approaches. This shift is reshaping how vehicles are designed, built, and operated, and at the same time, adds complexity to the production and running of an already very complex artifact. As software becomes a core driver of innovation, it also introduces a host of new challenges, ranging from regulatory compliance and cybersecurity to testing and verification of vehicle designs and operations. 
Introducing ever-growing software is essential to meet new laws and regulations (e.g., regarding emissions) and strengthen the market position.  

Nowadays, most automotive innovations stem from software developments, leading to the creation of complex embedded systems. These systems, also known as cyber-physical systems (CPSs), typically include at least a hundred Electronic Control Units (ECUs), each an independent CPS on its own, interconnected through bus systems with functions distributed across multiple units~\cite{Broy2007Engineering, Lami2016, Grimm2003, Chakraborty2016, Sundmark2011}. These innovations result in large-scale functions that can correspond to over 6,000 requirements~\cite{Juhke2020}. 
The consequential rise of complexity in vehicles to accommodate new functions and technologies drives outsourcing in the automotive industry and, thus, increases software development costs~\cite{Awedikian, Shaot2021}. Additionally, the vast number of variants, which can exceed 10 million, necessitates consistency~\cite{Guissouma}.

Much of the costs can be allocated to testing processes~\cite{Kushwaha2008, Ramler2006, Broy2006}. 
The automotive industry has adopted various testing methods and regulations (e.g., ISO 26262 and ISO 9001) in response to these challenges. Standards such as AUTOSAR~\cite{autosar} and ASPICE~\cite{aspice} try to address the complexities of modern vehicle technology and ensure the highest standards of quality and safety by offering software abstraction layers and standardized interfaces. For instance, AUTOSAR consists of two platforms, Classic and Adaptive, with very different approaches when it comes to testing. Whereas AUTOSAR Classic relies on tests performed during development, AUTOSAR Adaptive introduces, in addition, runtime verification through continuous monitoring and adaptation~\cite{Stawski}. The existence of both platforms operating simultaneously in vehicle systems makes testing even more challenging.



In ASPICE, especially in the Software Unit Verification Process, an identified weakness is the poorly defined test strategies (e.g., no explicit definition of the test scope, lack of test plans, or methods for test cases) concerning software unit verification activities, including static code verification, and unit testing~\cite{Falcini,Lachmann2013, Juhke2020}. 
Further, the execution of the software unit verification tests is performed quite informally with little evidence, like test case specifications,  a verification log, or a verification report. There is a heavy reliance on the verifier's experience without clear targets and procedures. 
Hence, even with the current established standards AUTOSAR and ASPICE, practical shortcomings persist: a seamless chain of methods and tools to support the entire life cycle of an ECU and their integration, from requirement specification to integration and testing, is lacking~\cite{Grimm2003, Broy2006, Juhke2020}. Emerging technologies are addressing these limitations. For instance, Shaout and Pattela highlight that up to 80\% of automotive software can now be automatically generated from models~\cite{Shaot2021}.


Furthermore, the challenges are not limited to a single Original Equipment Manufacturer (OEM). A vehicle is developed by a complex network of multiple tiers of suppliers that must be coordinated across different teams, companies, countries, and continents. Integrating multiple components and systems demands consistency, availability, and reliability in testing, especially on a global scale~\cite{Zyberaj2023, Grimm2003, Pretschner2007, Rajkumar2010}. Moreover, the outsourcing of testing-related tasks to suppliers is increasing, further complicating the testing process~\cite{Juhke2020, Sztipanovits2012}. As a result, the test landscape is highly context-specific, with substantial variation between OEMs, tier-1 suppliers, and tool vendors, as well as across product lines and markets.

\begin{figure}
    \centering
    \includegraphics[width=1\linewidth]{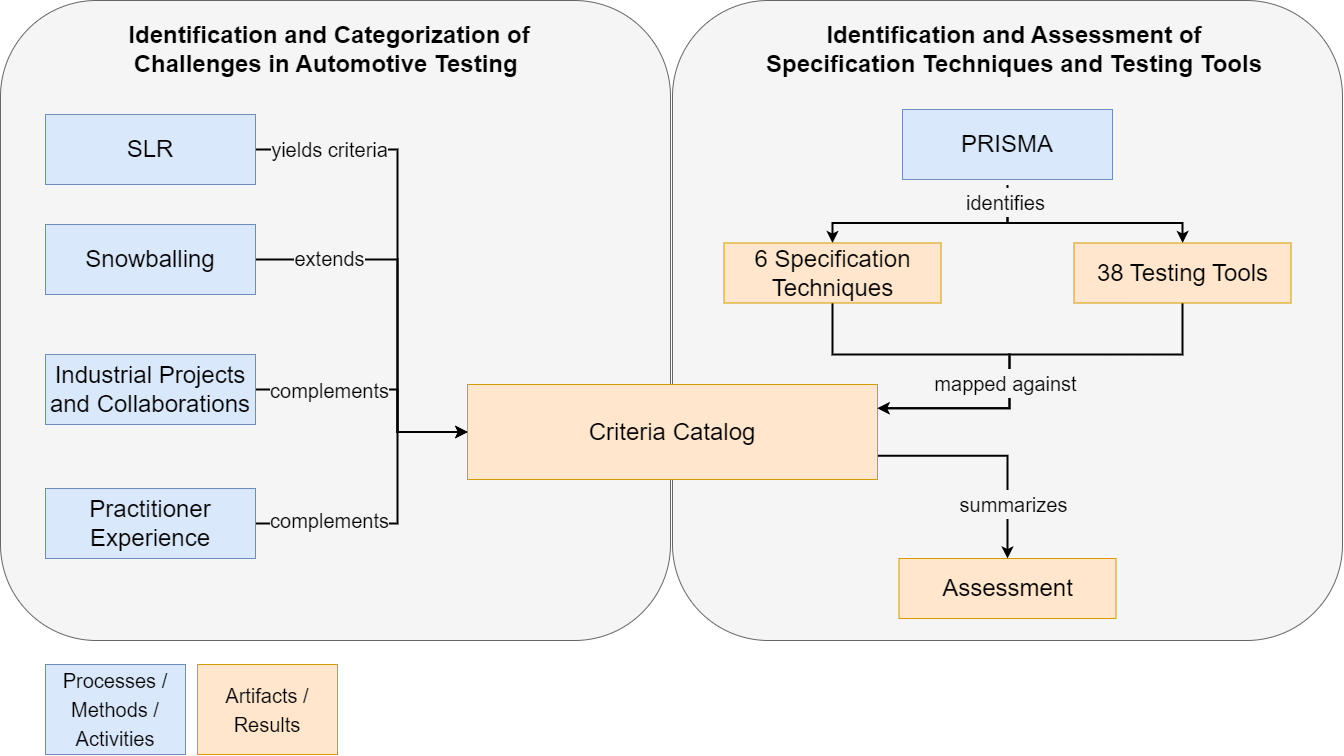}
 \caption{Overview of the research design: challenges and requirements are consolidated into a criteria catalog (left), which is then used to assess test case specification techniques and system testing tools identified via PRISMA (right).}
    \label{fig:overview}
\end{figure}

This article investigates the challenges, specific requirements, and tools currently employed in writing, running, and automating tests within the automotive industry. A fundamental step in our research is conducting a Systematic Literature Review (SLR)~\cite{Keele2007}, emphasizing the broader testing field with a particular focus on automotive applications. We identify further primary studies using the snowballing method~\cite{Wohlin2014}, providing a solid foundation to identify critical requirements and develop a comprehensive testing criteria catalog.

Following the Preferred Reporting Items for Systematic Reviews and Meta-Analyses (PRISMA) framework~\cite{Liberati2009}, we extensively search to identify relevant test specifications and tools from industrial practice and academic research. These resources are evaluated based on our criteria catalog, highlighting their applicability and potential gaps within the automotive domain. In summary, the contributions presented comprise:
\begin{enumerate}
	\item Deriving comprehensive requirements based on a SLR, practical experience, and insights from industry collaborations, forming the criteria catalog's foundation.
	\item Identifying and categorizing test case specification techniques and testing tools used in both industry and academia. 
	\item Assessing these specifications and tools against the identified requirements, thus pinpointing current strengths and significant gaps that remain to be addressed.
\end{enumerate}
Several existing surveys and frameworks cover adjacent aspects of our problem space. 
At the requirements level, Deckers and Lago survey domain-oriented specification techniques and Darif et al. review controlled natural languages for requirements specification; however, neither target automotive test case specifications or system testing tools \cite{Deckers2022,Darif2025}. 
At the testing level, surveys of automotive testing methodologies and AI-based techniques, along with evaluation frameworks for automated testing tools in agile projects, provide either high-level process overviews or criteria focused on specific technique families or development settings \cite{Laaroussi2024,Moseh2024}. Closer to our work, Juhnke et al. identify challenge categories for automotive test case specifications based on interviews at an OEM and several suppliers \cite{Juhnke2018}. 
However, to the best of our knowledge, none of these studies jointly (i) synthesizes automotive testing challenges across the life cycle, (ii) derives an explicit, prioritized criteria catalog from those challenges, and (iii) systematically maps concrete test case specification techniques and system testing tools against this catalog, which is the gap addressed by our study.

Figure~\ref{fig:overview} summarizes this two-stage design. The left-hand side shows how the SLR, snowballing, and industrial experience yield a criteria catalog that captures recurring challenges and improvement requirements. The right-hand side shows how we then apply a PRISMA-guided search to identify concrete test case specification techniques and system testing tools and evaluate them against this catalog. Since many of the synthesized challenges concern how tests are specified and how they are executed and maintained, we focus this second stage on these two layers as the most critical leverage points identified in the SLR.

In line with standard SLR practice, we focus on peer-reviewed publications and complement them with practitioner experience and state-of-practice knowledge from industrial projects, which informs both the criteria catalog and the mapping of techniques and tools. The significance of our research lies in systematically presenting existing automotive testing approaches and standards while identifying areas requiring further research and development. This work offers guidance toward more coherent and integrated testing strategies in heterogeneous, legacy-constrained environments, ultimately enhancing both theoretical research and industrial practice.
While we briefly discuss related domains such as artificial intelligence (AI) enhanced testing, the Internet of Things (IoT), vehicle-to-everything (V2X), and autonomous driving, a full systematic review of these areas is beyond the scope of this work and would require dedicated studies.

Section~\ref{sec:slr} describes our SLR methodology and derives a criteria catalog. Section~\ref{sec:specs_tools} applies PRISMA to identify test case specification techniques and tools and maps them to the catalog. The threats to validity are discussed in Section~\ref{threats}. Section~\ref{sec:rq-answers} synthesizes the results and answers the research questions. Section~\ref{sec:conclusion} concludes and~\ref{sec:background} summarizes key automotive standards.

\section{Systematic Literature Review (SLR)}
\label{sec:slr}
In Section~\ref{sec:method}, we introduce the SLR method and describe the results in Section~\ref{sec:soa}.

\subsection{Method}
\label{sec:method}
To gather and analyze relevant literature, we perform an SLR as defined in~\cite{Keele2007} and adopt the snowballing method~\cite{Wohlin2014}. We define the research questions using the PICOC (Population, Intervention, Comparison, Outcome, Context) framework, a common practice in evidence-based software engineering. Table~\ref{tab:rq_analysis} summarizes the PICOC dimensions and their mapping to our research questions, clarifying the scope and focus of our analysis.

\begin{description}
    \item[RQ1.] Which testing‑related challenges are reported for automotive software and cyber‑physical systems across the full test life cycle?
    \item[RQ2.] What technical and process requirements have been proposed to improve the efficiency and effectiveness of automotive system testing?
\end{description}

\begin{table}[htbp]
  \centering
  \renewcommand{\arraystretch}{1.25}
  \caption{Research questions mapped to PICOC dimensions}
  \label{tab:rq_analysis}
  \begin{tabularx}{\textwidth}{>{\bfseries}p{0.18\textwidth} X X}
    \toprule
    \rowcolor{gray!5}
    Dimension & RQ1 & RQ2 \\
    \midrule
    Population 
      & \multicolumn{2}{p{0.74\textwidth}}{\centering Automotive software / cyber-physical systems}  \\
    \rowcolor{gray!5}
    Intervention 
      & \textit{None} (observational) 
      & Requirements (process, tooling, metrics) \\
    Comparison 
      & \multicolumn{2}{p{0.74\textwidth}}{\centering Not applicable} \\ 
    \rowcolor{gray!5}
    Outcome 
      & Catalog and taxonomy of reported challenges 
      & Catalog of requirements that improve efficiency and effectiveness, with links to addressed challenges \\
    Context 
      & \multicolumn{2}{p{0.74\textwidth}}{\centering Any test level (unit to acceptance, including XiL and in-service) within the automotive domain} \\
    \bottomrule
  \end{tabularx}
\end{table}

We defined a controlled vocabulary based on domain knowledge and literature to maintain consistency and ensure comprehensiveness across database searches. Table~\ref{tab:control_voc} summarizes these terms grouped into semantic categories used for query construction, enhancing search precision and consistency. 

\begin{table}[htbp]
    \centering
    \caption{Controlled vocabulary for search strings}
    \label{tab:control_voc}
    \renewcommand{\arraystretch}{1.3}
    \begin{tabularx}{\textwidth}{>{\bfseries}p{0.25\textwidth} X}
        \toprule
        \rowcolor{gray!5}
         \textbf{Category} & \textbf{Controlled vocabulary} \\
        \midrule
        AUTOMOTIVE 
          & (automotive OR vehicle* OR ``road~vehicle'' OR ECU OR ``electronic~control~unit'' OR ``in~vehicle'') \\
        \rowcolor{gray!5}
        TEST 
          & (test* OR verification OR validation* OR validat* OR ``V\&V'' OR ``quality~assurance'') \\
        CHALLENGE 
          & (challenge* OR issue* OR problem* OR bottleneck* OR obstacle*) \\
        \rowcolor{gray!5}
        REQUIRE 
          & (requirement* OR need* OR criteri*) \\
        IMPROVE 
          & (optim* OR optimis* OR improv* OR enhanc* OR increas*) \\
        \bottomrule
    \end{tabularx}
\end{table}

The SMALL‑CAPS tokens in all queries refer to the controlled‑vocabulary blocks of Table~\ref{tab:control_voc}. Based on the research questions, the targeted generic master strings are: 
\begin{description}
  \item[SQ1 (Challenges)] {\small
        \textsc{challenge} \,W/3\,
        (\textsc{automotive\_title} \textbf{OR} \textsc{automotive\_abstract})
        \textbf{AND} \textsc{test}}

  \item[SQ2 (Requirements)] {\small
        \textsc{require} \textbf{AND} \textsc{improve}
        \textbf{AND} \textsc{automotive} \textbf{AND} \textsc{test}}
\end{description}

\noindent
The generic form omits field tags; database‑specific field scoping and proximity operators are shown in Tables \ref{tab:sting_database_1}–\ref{tab:sting_database_2}.`W/3` is replaced by `NEAR/3` for IEEE Xplore and ACM DL.

\textbf{SQ1} is composed of three logical blocks. The first block, CHALLENGE, contains testing-related problem terms searched specifically within titles to maximize topical relevance. The second block consists of AUTOMOTIVE domain terms appearing either in the title or abstract; these terms are combined with proximity operators (W/3 or NEAR/3) to ensure they closely relate to the challenge terms. The third block, TEST, filters explicitly for testing and verification vocabulary at the abstract level, retrieving only papers focused on testing contexts.

\textbf{SQ2} builds upon four logical blocks without a specific order. The REQUIRE block captures requirement-related nouns, while the IMPROVE block includes verbs related to optimization or enhancement. The AUTOMOTIVE block broadly searches domain-specific terminology across titles and abstracts. Finally, the TEST block focuses on testing and verification vocabulary. No proximity operators are used since SQ2 aims to identify any co-occurrence among these concepts. This approach maximizes the recall of literature discussing process or tool improvements within automotive system testing.

To effectively answer these research questions, we developed targeted search queries aligned with each RQ using Polyglot Search Translator~\cite{polyglot2023}, IEEE Syntax Helper~\cite{ieeesyntax2023}, and Search Strategy Translator~\cite{searchtranslator2024}. These queries leverage Boolean logic to ensure comprehensiveness and precision in identifying relevant literature, as detailed in Table~\ref{tab:sting_database_1} and Table~\ref{tab:sting_database_2}. The database‑specific syntax variants and the full RIS export are provided in the replication package.

The searches are performed across four of the best-known databases: (1) ACM Digital Library (DL), (2) IEEE Xplore Digital Library (DL), (3) Scopus, and (4) SpringerLink. This combination balances breadth (two multidisciplinary indices) with depth (three subject‑specific digital libraries). In an initial trial phase, we also queried Web of Science. However, after deduplication against ACM DL, IEEE Xplore, Scopus, and SpringerLink, Web of Science did not contribute additional eligible primary studies. To avoid redundant screening effort, we therefore based the final search protocol on the four reported databases.

The screening and data extraction were performed independently by two authors. In the first pass, both authors screened the same pilot set of records to calibrate inclusion and exclusion decisions. Disagreements were discussed in consensus meetings until a shared interpretation of the criteria was reached, and the decision rules were refined accordingly. The remaining records were then split between the authors for screening, with borderline cases flagged and jointly resolved. For data extraction and coding of challenges and requirements, both authors extracted information into a shared coding sheet. A pilot subset was double-coded and discussed to align category boundaries; the remaining papers were coded by one author and spot-checked by the other. We relied on iterative calibration and consensus to ensure consistent coding.

Publications are selected by applying online build-in filters based on the following criteria: (i) published between 2010 and 2025 to ensure the inclusion of the most recent studies, (ii) written in English, (iii) types of documents: journal articles, conference papers, workshop papers, survey papers, and books (iv) relevant to the automotive domain. 
For example, we included studies that primarily targeted embedded or cyber-physical systems but contained a dedicated automotive case study with explicit discussion of the test process, since they satisfied the domain relevance and peer-review criteria. In contrast, we excluded papers that proposed generic testing approaches or tools and mentioned automotive systems only in passing, without any domain-specific evaluation. Borderline records were discussed between the two reviewers; in cases of doubt, they were retained during the abstract screening process, and a final inclusion decision was made after reviewing the full text.

We chose 2010 as the lower bound to focus on the period in which modern automotive safety and architecture standards (e.g., ISO 26262 and AUTOSAR 3.x and later) and their associated testing toolchains became widely adopted, so that the synthesis reflects contemporary practice rather than pre-standardization methods. Backward snowballing occasionally led to the inclusion of seminal pre-2010 work that is still actively cited in recent automotive testing studies; the oldest paper in our corpus dates from 2008.

Exclusions were made for publications that were not accessible or were not peer-reviewed. The inclusion criteria were selected to ensure methodological rigor, recentness, and applicability to automotive system testing practices.

\begingroup
\footnotesize
\rowcolors{2}{white}{gray!5}
\setlength\LTpre{0pt}
\setlength\LTpost{0pt}
\begin{longtable}{
@{}
>{\bfseries}p{0.14\textwidth}
p{0.80\textwidth}
@{}
}
\caption{Database-specific queries for SQ1}
\label{tab:sting_database_1} \\
\toprule
\textbf{Database} & \textbf{Search Query} \\
\midrule
\endfirsthead

\multicolumn{2}{@{}l}{\small\slshape Table \thetable\ (continued)} \\[-3pt]
\toprule
\textbf{Database} & \textbf{Search Query} \\
\midrule
\endhead

\bottomrule
\endlastfoot

ACM DL &
title:(challenge* OR issue* OR problem* OR bottleneck* OR obstacle*) NEAR/3 ( title:(automotive OR vehicle* OR "road vehicle" OR "in vehicle") OR abstract:(automotive OR vehicle* OR "road vehicle" OR "in vehicle") ) AND abstract:(test* OR verification OR validation OR "V\&V" OR "quality assurance")\\

IEEE Xplore\- DL &
("Document Title":challenge* OR "Document Title":issue* OR "Document Title":problem* OR "Document Title":bottleneck* OR "Document Title":obstacle*) NEAR/3 ("Document Title":automotive OR "Document Title":vehicle* OR "Document Title":"road vehicle" OR "Document Title":"in vehicle") AND ("Abstract":test* OR "Abstract":verification OR "Abstract":validation OR "Abstract":"V\&V" OR "Abstract":"quality assurance") \\

Scopus &
TITLE-ABS-KEY( (challenge* OR issue* OR problem* OR bottleneck* OR obstacle*) W/3 (automotive OR vehicle* OR "road vehicle" OR "in vehicle") ) AND TITLE-ABS-KEY(test* OR verification OR validation OR "V\&V" OR "quality assurance") AND (PUBYEAR > 2009 AND PUBYEAR < 2026) \\

Springer\-Link &
(challenge* OR issue* OR problem* OR bottleneck* OR obstacle*) AND (automotive OR vehicle* OR "road vehicle" OR "in vehicle") AND (test* OR verification OR validation OR "V\&V" OR "quality assurance") \\
\end{longtable}
\endgroup

\begingroup
\footnotesize
\rowcolors{2}{white}{gray!5}
\setlength\LTpre{0pt}
\setlength\LTpost{0pt}
\begin{longtable}{
@{}
>{\bfseries}p{0.14\textwidth}
p{0.80\textwidth}
@{}
}
\caption{Database-specific queries for SQ2} 
\label{tab:sting_database_2} \\
\toprule
\textbf{Database} & \textbf{Search Query} \\
\midrule
\endfirsthead

\multicolumn{2}{@{}l}{\small\slshape Table \thetable\ (continued)} \\[-3pt]
\toprule
\textbf{Database} & \textbf{Search Query} \\
\midrule
\endhead

\bottomrule
\endlastfoot

ACM DL &
Title:(automotive OR vehicle* OR ADAS OR "in-vehicle") AND
Abstract:(automotive OR vehicle* OR ADAS OR "in-vehicle") AND
Title:(test* OR verification OR validation* OR "model-based testing" OR MBT OR MBSE OR
"continuous integration" OR CI OR "continuous delivery" OR CD OR regression OR coverage) AND
Abstract:(test* OR verification OR validation* OR "model-based testing" OR MBT OR MBSE OR
"continuous integration" OR CI OR "continuous delivery" OR CD OR regression OR coverage) AND
AllField:("in-the-loop" OR "hardware-in-the-loop" OR "software-in-the-loop" OR
"vehicle-in-the-loop" OR HIL OR SIL OR MIL OR VIL OR XIL) AND
AllField:(requirement* OR need* OR criteri*) AND
AllField:(optim* OR improv* OR enhanc* OR increas* OR efficien* OR effectiv* OR
matur* OR scalab* OR automati*) \\

IEEE Xplore DL &
(("Document Title":automotive OR "Document Title":vehicle* OR "Document Title":ADAS OR "Document Title":"in‑vehicle")
AND ("Abstract":automotive OR "Abstract":vehicle* OR "Abstract":ADAS OR "Abstract":"in‑vehicle")
AND ("Document Title":test OR "Document Title":verification OR "Document Title":validation OR
"Document Title":"model‑based testing" OR "Document Title":"continuous integration" OR
"Document Title":"continuous delivery" OR "Document Title":regression OR "Document Title":coverage)
AND ("Abstract":test OR "Abstract":verification OR "Abstract":validation OR
"Abstract":"model‑based testing" OR "Abstract":"continuous integration" OR
"Abstract":"continuous delivery" OR "Abstract":regression OR "Abstract":coverage)
AND ("All Metadata":"in‑the‑loop" OR "All Metadata":"hardware‑in‑the‑loop" OR
"All Metadata":"software‑in‑the‑loop" OR "All Metadata":"vehicle‑in‑the‑loop" OR
"All Metadata":HIL OR "All Metadata":SIL OR "All Metadata":MIL OR "All Metadata":VIL OR "All Metadata":XIL)
AND ("All Metadata":requirement* OR "All Metadata":need* OR "All Metadata":criteri*)
AND ("All Metadata":optim* OR "All Metadata":improv* OR "All Metadata":efficien* OR
"All Metadata":effectiv* OR "All Metadata":automati*)) \\

Scopus &
TITLE-ABS-KEY (
( automotive OR ADAS OR "in-vehicle" )
AND ( "software test*" OR "system test*" OR "model-based testing" OR MBT OR MBSE
OR "software verification" OR "software validation"
OR "continuous integration" OR "CI/CD" OR "CI pipeline"
OR regression OR "scenario-based testing" )
AND ( "hardware-in-the-loop" OR "software-in-the-loop"
OR "vehicle-in-the-loop" )
AND ( requirement* OR specification* )
AND ( optim* OR improv* OR efficien* OR automat* )
)
AND PUBYEAR > 2009 AND PUBYEAR < 2026
AND LIMIT-TO ( SUBJAREA , "ENGI" ) AND LIMIT-TO ( SUBJAREA , "COMP" )
AND LIMIT-TO ( LANGUAGE , "English" )
AND NOT TITLE-ABS-KEY ( "sensor coverage" OR battery ) \\

Springer\-Link &
((automotive OR vehicle* OR adas OR "in-vehicle")
AND (test* OR verification OR validation OR "model-based testing")
AND ("in-the-loop" OR HIL OR SIL OR MIL OR VIL OR XIL)
AND (optimisation OR improve OR efficiency OR effective OR automation)
AND ("test case" OR checklist OR requirement OR criteria)) \\

\end{longtable}
\endgroup

\begin{figure}
    \centering
    \includegraphics[width=1\linewidth]{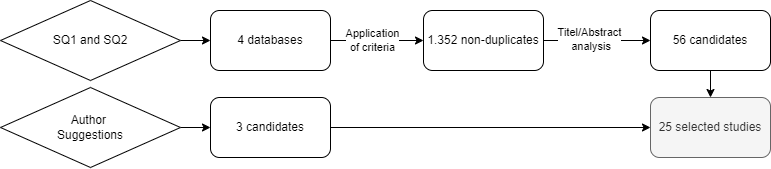}
    \caption{Flowchart of the Systematic Literature Review (SLR): key steps from search queries to the final set of selected studies.}
    \label{fig:slr}
\end{figure}

As shown in Figure~\ref{fig:slr}, the search initially identified 1.617 records from which we removed 265 duplicates. The next phase involved filtering these publications by title and abstract, resulting in 56 articles suitable for full-text analysis. Furthermore, we included 3 candidates suggested by the authors. From these, we identified 25 papers that qualified as primary studies (PSs). 
\begingroup
\footnotesize
\rowcolors{2}{white}{gray!5}
\setlength\LTpre{0pt}
\setlength\LTpost{0pt}
\begin{longtable}{
@{}
>{\centering\arraybackslash}p{0.9cm}    
>{\raggedright\arraybackslash}p{2.2cm}  
>{\raggedright\arraybackslash}p{5.5cm}  
>{\centering\arraybackslash}p{0.9cm}    
>{\centering\arraybackslash}p{1.6cm}    
>{\centering\arraybackslash}p{1.2cm}    
@{}
}
\caption{Comprehensive overview of primary studies identified in the Systematic Literature Review (SLR)} 
\label{tab:slr_ps} \\
\toprule
\textbf{PS ID} & \textbf{First\- Author} & \textbf{Title} & \textbf{Year} & \textbf{Publisher} & \textbf{Ref.} \\
\midrule
\endfirsthead

\multicolumn{6}{@{}l}{\small\slshape Table \thetable\ (continued)}\\[-4pt]
\toprule
\rowcolor{gray!5}
\textbf{PS ID} & \textbf{First\- Author} & \textbf{Title} & \textbf{Year} & \textbf{Publisher} & \textbf{Ref.} \\
\midrule
\endhead

\bottomrule
\endlastfoot
\addlinespace[2pt]
PS1 & Altinger, Harald 
    & Testing methods used in the automotive industry: results from a survey 
    & 2014 
    & ACM 
    & ~\cite{Altinger2014}\\
\addlinespace[2pt]
PS2 & Amarnath,\newline Rakshith 
    & Dependability Challenges in the Model-Driven Engineering of Automotive Systems   
    & 2016 
    & IEEE 
    & ~\cite{Amarnath2016}\\
\addlinespace[2pt]
PS3 & Arrieta, Aitor 
    & Employing Multi-Objective Search to Enhance Reactive Test Case Generation and Prioritization for Testing
  Industrial Cyber-Physical Systems 
    & 2018 
    & IEEE 
    & ~\cite{Arrieta2018}\\
\addlinespace[2pt]
PS4 & Atoum, Issa 
    & Challenges of Software Requirements Quality Assurance and Validation: A Systematic Literature Review 
    & 2021  
    & IEEE 
    & ~\cite{Atoum2021}\\
\addlinespace[2pt]
PS5 & Bashroush,\newline Rabih 
    & CASE Tool Support for Variability Management in Software Product Lines 
    & 2017 
    & ACM 
    & ~\cite{Bashroush2017}\\
\addlinespace[2pt]
PS6 & Berger, Thorsten 
    & A survey of variability modeling in industrial practice 
    & 2013 
    & ACM 
    & ~\cite{Berger2013}\\
\addlinespace[2pt]
PS7 & Bringmann, Eckard 
    & Model-Based Testing of Automotive Systems 
    & 2008 
    & IEEE 
    & ~\cite{Bringmann2008}\\
\addlinespace[2pt]
PS8 & Broy, Manfred 
    & Seamless model-based development: From isolated tools to integrated model engineering environments & 2010 
    & IEEE 
    & ~\cite{Broy2010}\\
\addlinespace[2pt]
PS9 & Bruel,\newline Jean-Michel 
    & The Role of Formalism in System Requirements 
    & 2021 
    & ACM 
    & ~\cite{bruel2020}\\
\addlinespace[2pt]
PS10 & Cederbladh,\newline Johan 
     & Early Validation and Verification of System Behaviour in Model-based Systems Engineering: A Systematic Literature Review 
     & 2024 
     & ACM 
     & ~\cite{Cederbladh2024}\\
\addlinespace[2pt]
PS11 & D'Ambrosio, Joseph 
     & Systems engineering challenges and MBSE opportunities for automotive system design 
     & 2017 
     & IEEE 
     & ~\cite{Ambrosio2017}\\
\addlinespace[2pt]
PS12 & Derler, Patricia 
     & Modeling Cyber–Physical Systems 
     & 2012 
     & IEEE 
     & ~\cite{Derler2012}\\
\addlinespace[2pt]
PS13 & Feiler, Peter 
     & Model-based engineering with AADL: An introduction to the SAE architecture analysis design language 
     & 2012 
     & Pearson International 
     & ~\cite{Feiler2012}\\
\addlinespace[2pt]
PS14 & Garousi, Vahid 
     & What industry wants from academia in software testing?: Hearing practitioners' opinions 
     & 2017 
     & ACM 
     & ~\cite{Garousi2017}\\
\addlinespace[2pt]
PS15 & Goswami, Dip 
     & Challenges in automotive cyber-physical systems design 
     & 2012 
     & IEEE 
     & ~\cite{Goswami2012}\\
\addlinespace[2pt]
PS16 & Greca, Renan 
     & State of Practical Applicability of Regression Testing Research: A Live Systematic Literature Review 
     & 2023 
     & ACM 
     & ~\cite{Greca2023}\\
\addlinespace[2pt]
PS17 & Gruber, Kristina 
     & Integrated Description of Functional and Non-Functional Requirements for Automotive Systems Design Using SysML 
     & 2017 
     & IEEE 
     & ~\cite{Gruber2017}\\
\addlinespace[2pt]
PS18 & Habibullah,\newline Kahn\newline Mohammad 
     & Requirements and software engineering for automotive perception systems: an interview study 
     & 2024 
     & Springer 
     & ~\cite{Habibullah2024}\\
\addlinespace[2pt]
PS19 & Hagar, John D. 
     & IoT System Testing 
     & 2022 
     & Apress 
     & ~\cite{Hagar2022}\\
\addlinespace[2pt]
PS20 & Haghighatkhah, Alireza 
     & Improving the State of Automotive Software Engineering 
     & 2017 
     & IEEE 
     & ~\cite{Haghighatkhah2017}\\
\addlinespace[2pt]
PS21 & Imke, Drave 
     & SMArDT modeling for automotive software testing 
     & 2018 
     & wiley 
     & ~\cite{Drave2018}\\
\addlinespace[2pt]
PS22 & Juhnke,\newline Katharina 
     & A Quality Model and Checklists for Reviewing Automotive Test Case Specifications 
     & 2022 
     & Springer
     & ~\cite{Juhnke2022}\\
\addlinespace[2pt]
PS23 & Juhnke,\newline Katharina 
     & Challenges concerning test case specifications in automotive software testing: assessment of frequency and criticality 
     & 2020 
     & Springer 
     & ~\cite{Juhke2020}\\
\addlinespace[2pt]
PS24 & Karhapää, Pretti 
     & Strategies to manage quality requirements in agile software development: a multiple case study
     & 2021 
     & Springer 
     & ~\cite{Karhapää2021}\\
\addlinespace[2pt]
PS25 & Kasoju,\newline Abhinaya 
     & Analyzing an automotive testing process with evidence-based software engineering 
     & 2013 
     & Elsevier 
     & ~\cite{Kasoju2013}\\
\addlinespace[2pt]
PS26 & Knauss, Eric 
     & Continuous Integration Beyond the Team: A Tooling Perspective on Challenges in the Automotive Industry 
     & 2016 
     & ACM 
     & ~\cite{Knauss2016}\\
\addlinespace[2pt]
PS27 & Kranabitl, Philipp 
     & A fundamental concept for linking methods, system models, and specific models: Future of systems
  engineering
     & 2024 
     & Springer 
     & ~\cite{Kranabitl2024}\\
\addlinespace[2pt]
PS28 & Lachmann, Remo 
     & Towards efficient and effective testing in automotive software development 
     & 2014 
     & Gesellschaft für Informatik e.V. 
     & ~\cite{Lachmann2014}\\
\addlinespace[2pt]
PS29 & Lahami, Mariam 
     & A survey on runtime testing of dynamically adaptable and distributed systems 
     & 2021 
     & Springer 
     & ~\cite{Lahami2021}\\
\addlinespace[2pt]
PS30 & Liao, Lei 
     & Vehicle Domain-Specific Language: Unifying Modeling and Code Generation for Low-Code Automotive Development 
     & 2024 
     & ACM 
     & ~\cite{Liao2024}\\
\addlinespace[2pt]
PS31 & Lu, Jinzhi 
     & A Service-Oriented Tool-Chain for Model-Based Systems Engineering of Aero-Engines 
     & 2018
     & IEEE
     & ~\cite{Lu2018}\\
\addlinespace[2pt]
PS32 & Masmoudi, Chedhli 
     & Adopting formal methods on requirements verification and validation for cyber-physical systems: A systematic literature review 
     & 2022 
     & Elsevier 
     & ~\cite{Masmoudi2022}\\
\addlinespace[2pt]
PS33 & Minani,\newline Jean Baptiste 
     & A Systematic Review of IoT Systems Testing: Objectives, Approaches, Tools, and Challenges 
     & 2024 
     & IEEE 
     & ~\cite{Minani2024}\\
\addlinespace[2pt]
PS34 & Mohd-Shafie, Muhammad Luqman
     & Model-based test case generation and prioritization: a systematic literature review 
     & 2022 
     & Springer 
     & ~\cite{Mohd-Shafie2022}\\
\addlinespace[2pt]
PS35 & Moukahal, Lama J. 
     & Vehicle Software Engineering (VSE): Research and Practice 
     & 2020 
     & IEEE 
     & ~\cite{Moukahal2020}\\
\addlinespace[2pt]
PS36 & Naimi, Lahbib 
     & A new approach for automatic test case generation from use case diagram using LLMs and prompt engineering 
     & 2024 
     & IEEE 
     & ~\cite{Naimi2024}\\
\addlinespace[2pt]
PS37 & Norheim,\newline Johannes J. 
     & Challenges in applying large language models to requirements engineering tasks 
     & 2024 
     & Cambridge University Press 
     & ~\cite{Norheim2024}\\
\addlinespace[2pt]
PS38 & Petrenko, Alexandre 
     & Model-based testing of automotive software: Some challenges and solutions 
     & 2015 
     & ACM, EDAC, IEEE 
     & ~\cite{Petrenko2015}\\
\addlinespace[2pt]
PS39 & Schäfer,\newline Andreas 
     & Variability realization in model-based system engineering using software product line techniques: an industrial perspective 
     & 2021  
     & ACM 
     & ~\cite{Schäfer2021}\\
\addlinespace[2pt]
PS40 & Schroeder, Jan 
     & Challenges from Integration Testing using Interconnected Hardware-in-the-Loop Test Rigs at an Automotive OEM: An Industrial Experience Report 
     & 2015 
     & ACM 
     & ~\cite{Schroeder2015}\\
\addlinespace[2pt]
PS41 & Sztipanovits, Janos 
     & Model and Tool Integration Platforms for Cyber–Physical System Design 
     & 2018
     & IEEE 
     & ~\cite{Sztipanovits2018}\\
\addlinespace[2pt]
PS42 & Tahir, Zaid 
     & Coverage based testing for VV and Safety Assurance of Self-driving Autonomous Vehicles: A Systematic Literature Review 
     & 2020 
     & IEEE 
     & ~\cite{Tahir2020}\\
\addlinespace[2pt]
PS43 & Tukur,\newline Muhammad 
     & Requirement engineering challenges: A systematic mapping study on the academic and the industrial perspective 
     & 2021 
     & Springer 
     & ~\cite{Tukur2021}\\
\addlinespace[2pt]
PS44 & Wang, Yuqing
     & Improving test automation maturity: A multivocal literature review 
     & 2022 
     & wiley 
     & ~\cite{Wang2022}\\
\addlinespace[2pt]
PS45 & Wiecher, Carsten 
     & Scenarios in the Loop: Integrated Requirements Analysis and Automotive System Validation 
     & 2020 
     & ACM, IEEE 
     & ~\cite{Wiecher2020}\\
\addlinespace[2pt]
PS46 & Yu, Huafeng 
     & The challenge of interoperability: model-based integration for automotive control software 
     & 2015 
     & ACM 
     & ~\cite{Yu2015}\\
\addlinespace[2pt]
PS47 & Zahid, Farzana 
     & A systematic mapping of semi-formal and formal methods in requirements engineering of industrial cyber-physical systems 
     & 2022 
     & Springer 
     & ~\cite{Zahid2022}\\
\addlinespace[2pt]
PS48 & Zampetti, Fiorella 
     & Continuous Integration and Delivery Practices for Cyber-Physical Systems: An Interview-Based Study 
     & 2023 
     & ACM 
     &~\cite{Zampetti2023}\\
\addlinespace[2pt]
PS49 & Zhao, Liping 
     & Natural Language Processing for Requirements Engineering: A Systematic Mapping Study 
     & 2022 
     & ACM 
     & ~\cite{Zhao2021}\\
\addlinespace[2pt]
PS50 & Zimmermann, Markus 
     & Formulating engineering systems requirements 
     & 2023 
     & Springer 
     & ~\cite{Zimmermann2023}    \\

\end{longtable}
\endgroup

The snowballing method starts with a backward search, reviewing all references from each PS, comprising 1,634 studies. We applied the same inclusion and exclusion criteria and removed 45 duplicates. After a comprehensive screening of the titles, abstracts, and full texts of the remaining 505 studies, we identified seven additional PSs.

In the subsequent forward snowballing phase, the 25 original PSs led to the identification of 621 citations. We evaluated these citations using the established screening process and selected 114 for further detailed screening, eventually adding ten new PSs. Table~\ref{tab:snowballing} summarizes the results of the snowballing process.

The second round of backward snowballing targeted the 17 studies identified in the initial round. After thoroughly reviewing their references, we retrieved 34 potential studies and obtained three additional PSs. Lastly, a second forward snowballing of these 16 studies yielded 65 potential studies, five of which met our criteria and were included as new PSs. After completing the snowballing method, we compiled 50 PSs. Table~\ref{tab:slr_ps} presents a detailed list of all PSs.

\begin{table}[h]
  \centering
  \footnotesize
  \caption{Backward and forward snowballing process used to complete the literature findings}
  \label{tab:snowballing}
  \begin{tabular}{@{} l c c c @{}}
    \toprule
     \rowcolor{gray!5}
    \textbf{Snowballing} & \textbf{Round} & \textbf{Articles Retrieved} & \textbf{Articles Included} \\
    \midrule

    Backward  & 1 & 505 & 7  \\
    Forward   & 1 & 114 & 10 \\

    Backward  & 2 &  34 & 3  \\
    Forward   & 2 &  65 & 5  \\
    \midrule

    \textbf{Total}  &  & \textbf{718} & \textbf{25} \\
    \bottomrule
  \end{tabular}
\end{table}

\begin{figure}[htbp]
  \centering
  \includegraphics[width=.95\linewidth]{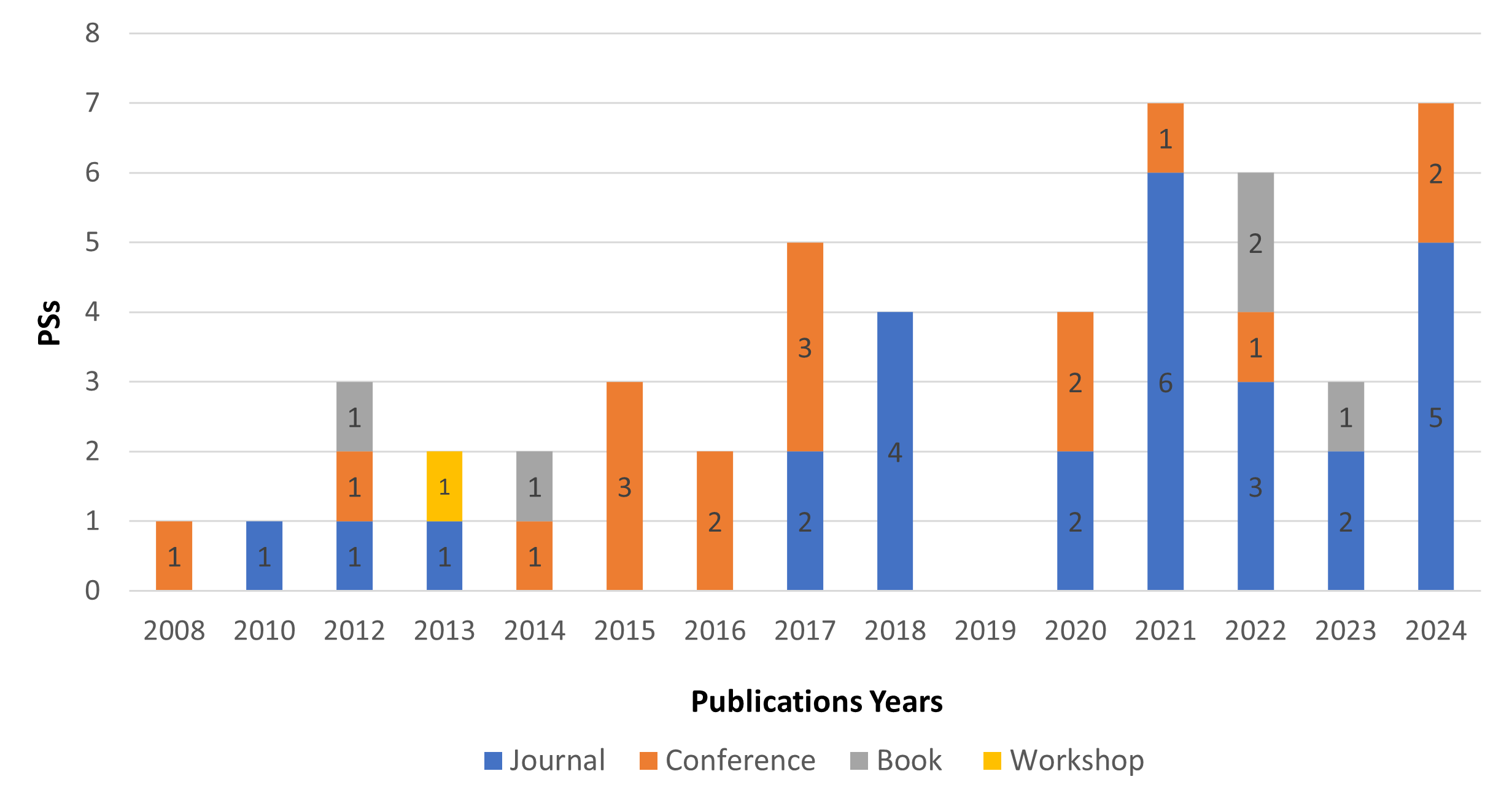}

  \smallskip     

  \textbf{(a)} Year-wise distribution of the 50 primary studies
  \par\smallskip  


  \includegraphics[width=.55\linewidth]{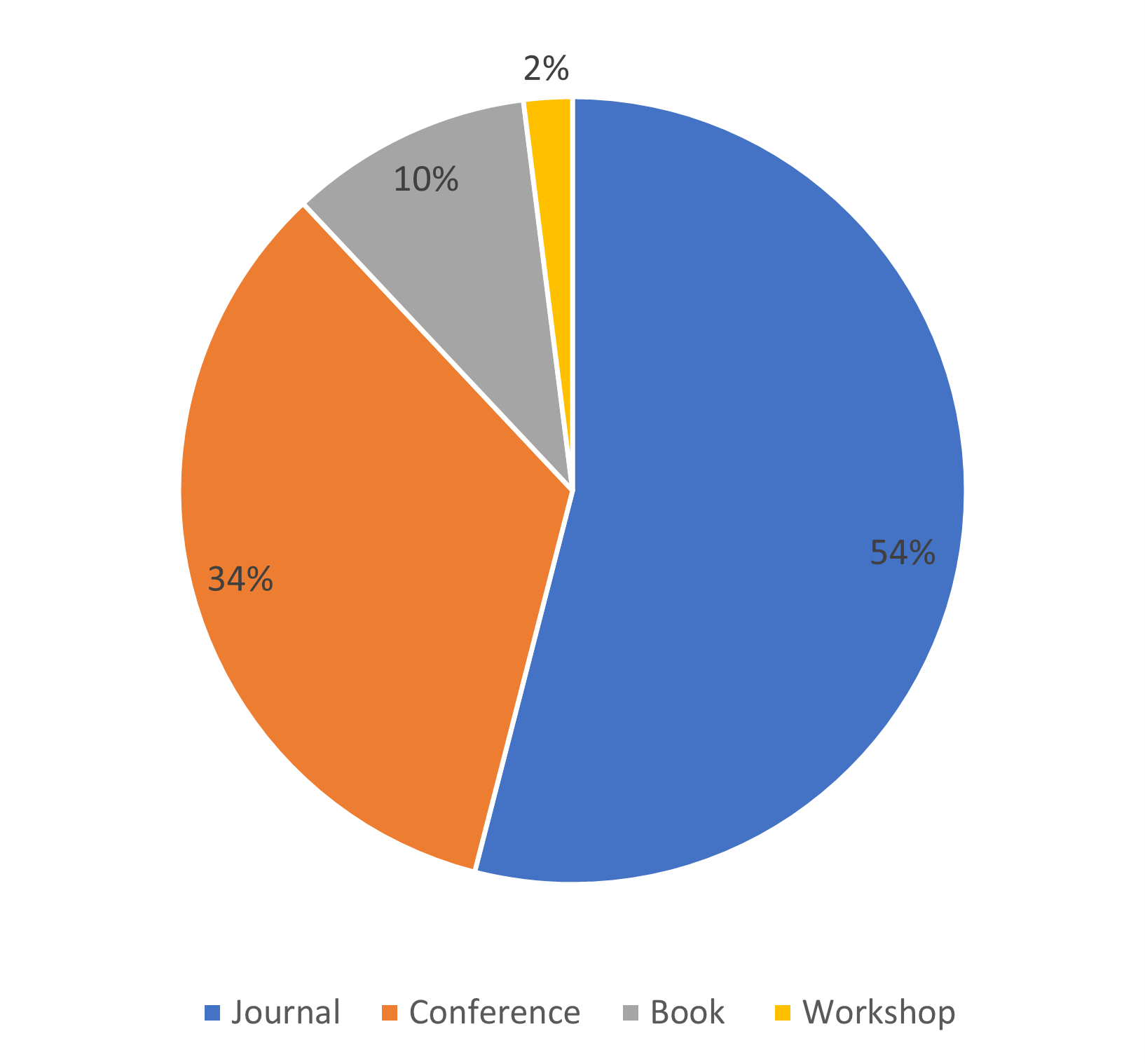}

  \smallskip
  \textbf{(b)} Primary studies by publication type

  \caption{Overview of primary studies by (a) publication year and
           (b) publication type.}
  \label{fig:overview_slr_year_type}
\end{figure}

Next, we extract the bibliometric characteristics of the compiled 50 PSs. Figure~\ref{fig:overview_slr_year_type} depicts the distribution of these studies by publication type from 2008 to 2024. 
%

We evaluated the authors' affiliations across the PSs and found that the majority (24 PSs) were affiliated with academic institutions (universities and research centers). In 20 cases, these academic authors collaborated with industry partners, including automotive OEMs such as Volvo, Audi, Daimler (now known as Mercedes-Benz), BMW, and General Motors, as well as suppliers such as Bosch. Industries facing similar challenges, such as aviation, participated in these collaborations. Notably, three PSs were conducted without any academic partner. This distribution indicates that the evidence base is skewed toward academic or joint academic–industry work, while large-scale, purely industrial deployments are less visible in the peer-reviewed literature.

Integrating academia and industry highlights a shared commitment to connecting theoretical research with practical, real-world applications. It demonstrates the direct relevance of these research challenges to current industry needs.

The countries affiliated with the authors predominantly include nations known for their traditional automakers.
Germany leads with 18 PSs, the USA with ten, and Sweden with six. Collaborations between industry and academia are most prevalent in Germany, accounting for nine PSs, Sweden with five, and the USA with four.

In summary, we conducted an SLR and snowballing method to identify relevant studies in automotive testing. Of the 50 PSs identified, most originate from countries with a well-established automotive industry, although not all studies focus exclusively on automotive topics. More than 50\% of these studies are purely academic, yet approximately 40\% involve collaborations between academia and industry partners.

\subsection{Results}
\label{sec:soa} 
\begin{figure}
    \centering
    \includegraphics[width=1\linewidth]{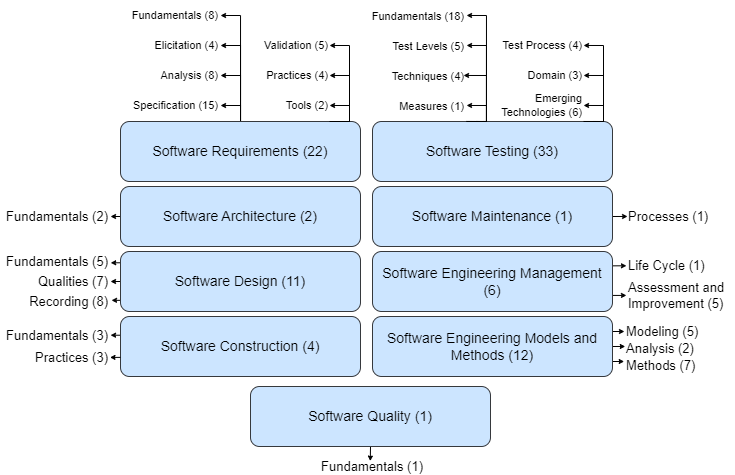}
    \caption{Classification of primary studies across nine SWEBOK Knowledge Areas (KAs)}
    \label{fig:swebok}
\end{figure}

The review of the 50 PSs is structured using the Guide to the Software Engineering Body of Knowledge v4.0 (SWEBOK)~\cite{SWEBOK2024}. 
We organize the findings by its Knowledge Areas (KAs) and derived subcategories. Within each group, we summarize recurring patterns rather than describing individual studies in detail. 
In the present case, we focus on nine KAs and derived subcategories that capture the dominant themes in the primary studies, as shown in Figure~\ref{fig:swebok}.

\subsubsection{Software Requirements (22 Studies)}

\textit{Software Requirements Fundamentals}. The automotive industry faces significant challenges in managing system requirements, particularly due to interoperability with legacy systems and the distributed nature of development. These challenges are exacerbated by communication gaps, cultural differences, and organizational inefficiencies within large projects, often stemming from distributed teams, hierarchical structures, and lack of tools~\cite{Habibullah2024, Karhapää2021, Masmoudi2022, Moukahal2020, Schroeder2015, Tukur2021, Zahid2022, Zimmermann2023}. Customers or clients often lack a clear understanding of system requirements, including the scope and specific functional and non-functional requirements. This lack of clarity makes it challenging for software engineers to develop solutions that accurately meet customer needs. Additionally, software engineers sometimes lack the necessary domain knowledge, limiting their ability to validate and accurately implement technical requirements, leading to suboptimal design choices. Furthermore, tight project schedules due to the pressure to reduce time-to-market can lead to a rushed development process. This often results in inadequate interactions between developers and customers, compromising a thorough understanding of requirements~\cite{Tukur2021}.\\\\
\textit{Requirements Elicitation}. Requirements and additional necessary information are distributed and must first be collected with several tools (e.g., IBM DOORS or Excel lists) and different stakeholders. The distribution challenge recurs in test cases where it aggravates the verification of the completeness of the test scope and encourages redundant executions on different test platforms~\cite{Juhke2020, Ambrosio2017}. Habibullah et al.\ and Tukur et al.\ also criticize the lack of a comprehensive requirement model and the unsystematic capture of requirements~\cite{Habibullah2024, Tukur2021}.\\\\
\textit{Requirements Analysis}. The primary triggers for most software failures are inconsistent or hidden requirements. The quality characteristics most frequently discussed are consistency (20.2\%), unambiguity (16.9\%), and completeness and correctness (15.7\% each). In contrast, understandability, reusability, unexpected dependencies, variability, and testability are not addressed~\cite{Atoum2021}. In the survey by Juhnke et al.~\cite{Juhke2020}, interviewees even report that the specifications are often not available in time, forcing testing personnel to design test cases based on outdated previous requirements and test with temporary placeholders for signals, which must be replaced later on and create a considerable post-processing effort. Furthermore, tracing between changing requirements and test specifications with a meaningful coverage metric seems an ongoing challenge in the industry~\cite{Garousi2017, Norheim2024, Tukur2021, Zimmermann2023}.\\\\
To address these challenges, defining requirements that comply with high standards is crucial. According to the ISO/IEC/IEEE 29148 standard, well-formed requirements should be unambiguous, correct, verifiable, complete, consistent, and feasible~\cite{Masmoudi2022}. Bruel et al.~\cite{bruel2020} established nine criteria to assess the suitability of requirements: (I) System vs. Environment (Scope), (II) Audience, (III) Level of Abstraction, (IV) Associated Method, (V) Traceability, (VI) Coverage, (VII) Semantic Definition, (VIII) Tool Support, and (IX) Verifiability.\\\\
\enlargethispage{-\baselineskip}
\textit{Requirements Specification}. Requirement specifications, essential for developing test cases, are categorized into five types: (I) natural language, (II) semi-formal, (III) automata or graphs, (IV) mathematical formalism, and (V) programming-language-based~\cite{bruel2020, Altinger2014, Norheim2024}. Predominantly, specifications are written in natural or semi-formal languages, which influences the quality of test case design, often leading to variability in quality due to under- or over-specified requirements~\cite{Garousi2017, Kasoju2013, Lachmann2014, bruel2020, Atoum2021, Zhao2021, Drave2018, Gruber2017, Habibullah2024, Masmoudi2022, Tukur2021, Yu2015, Zahid2022}. Zhao et al.\ address the ongoing challenges of using natural language in requirements engineering, noting that natural language is expected to remain the predominant medium for documenting requirements despite its limitations~\cite{Zhao2021}. They highlight a significant gap between laboratory experiments and the practical industrial application of Natural Language Processing for Requirements Engineering (NLP4RE) and show that only 13\%  of the linguistic analysis tools are available for industrial use after publication. Furthermore, challenges such as limited requirement-specific data, inconsistent data annotation, and inadequate definition of RE use cases further hinder the application of LLM-based NLP methods for RE.  The industry faces additional adoption barriers due to the lack of experience of RE professionals in Machine Learning (ML) and NLP and concerns about the imperfect performance of models in regulated or safety-critical domains~\cite{Norheim2024}.

Yu et al.\ recommend formal specifications that are incremental, composable, and reusable~\cite{Yu2015}. This approach facilitates model-based integration and formal verification, enabling specifications to remain independent of specific languages, tools, and platforms, leading to broader adaptation and reusability across the automotive industry.\\\\
\textit{Requirements Validation}. Several challenges can complicate the requirements validation process. Poor identification of stakeholders and their roles can hinder project success, as effective relationship management is crucial. Failure to meet stakeholders' needs and expectations also risks project failure if the end product does not satisfy their requirements. Long feedback cycles can delay requirements verification, potentially causing projects to miss critical deadlines. Furthermore, conflicts among stakeholders during requirements elicitation and analysis can hold back progress, emphasizing the importance of effective conflict detection and resolution strategies~\cite{Tukur2021, Zahid2022}. Verifying and validating multi-level system requirements are critical during two primary transformation phases: initially when translating stakeholder needs into system requirements, and subsequently when these requirements are transformed and allocated into subsystem requirements~\cite{Masmoudi2022}. Additionally, practical prototyping in combination with other techniques (e.g., knowledge-oriented approaches, formal models, or modeling) is helpful for stakeholders and developers. It provides early detection of requirement errors and prevents rework~\cite{Atoum2021, Masmoudi2022, Yu2015}.\\\\
\textit{Practical Considerations}. The traceability is aggravated by insufficient access to references referring to other specifications. This issue is observed in cooperation with suppliers and departments in the same company~\cite{Juhke2020, Haghighatkhah2017, Tukur2021, Zahid2022}. Furthermore, requirement engineers often fail to prioritize requirements based on factors such as criticality, importance, return on investment, and cost of delay. This lack of prioritization can lead to incomplete projects within the specified timeframe~\cite{Tukur2021}.\\\\
\textit{Software Requirements Tools}. Software requirement tools include various technologies, e.g., natural language processors (NLPs), machine learning (ML) algorithms, converters, assistants, and simulation tools. Requirements can be expressed in various formats, such as textual descriptions, use cases, class diagrams, and user stories. There is a lack of automated tools for the inspection process; 54.55\% of the evaluated requirement validation techniques do not mention a toolset~\cite{Atoum2021}. However, automatically and manually derived model-based test suites are shown to detect requirement errors better than handcrafted test suites~\cite{Haghighatkhah2017}.

\begin{keytakeaway}
\textbf{Key Takeaway:} The automotive industry faces significant challenges in software requirements management, including unclear specifications, poor traceability, reliance on natural language, and lack of automation in validation. Addressing these requires better tools, structured processes, stakeholder collaboration, and AI-driven automation.
\end{keytakeaway}

\subsubsection{Software Architecture (2 Studies)}

\textit{Software Architecture Description}. One of the main issues in the composition of models, particularly in automotive system design, is semantic interoperability, especially for timing specifications. The gap between software and architecture design often exacerbates timing concerns when the system is integrated. AUTOSAR is a well-established architecture framework in the automotive industry (see online supplementary material). Another standard, the Architecture Analysis and Design Language (AADL) by SAE International, focuses on modeling software-intensive systems, especially embedded, real-time, and safety-critical applications. AADL is recommended for use in the automotive, aerospace, and industrial automation industries. AADL facilitates static (design-time) and dynamic (runtime) modeling, defining system component behaviors and interactions. This approach enables early error detection, improves system integration through formal models, and ensures architectural consistency with well-defined interfaces~\cite{Feiler2012, Yu2015}.

\begin{keytakeaway}
\textbf{Key Takeaway:}  The automotive industry relies on standardized architecture frameworks such as AUTOSAR and AADL to address semantic interoperability, timing constraints, and a gap between software and architectural design. These frameworks enable early error detection, formal verification, and architectural consistency.
\end{keytakeaway}

\subsubsection{Software Design (11 Studies)}

\textit{Software Design Fundamentals}. The traditional automotive design flow is structured into two phases: (i) designing the control algorithm and (ii) implementing it on a specific embedded systems platform~\cite{Goswami2012}.

Software and hardware architectural aspects often impose restrictions on system design, yet these constraints are often neither explicitly nor formally specified. It is essential to integrate these aspects into a unified architectural view.
The automotive industry continues to separate functional models from software and hardware models. Software predominantly exists in source code form, while models are mainly utilized for documentation and argumentation purposes. This approach results in system designs that rely heavily on implicit expert knowledge~\cite{Amarnath2016, Ambrosio2017}.
Although there is a shift towards integrating architectural views using tools like SysML, AADL, or PREEVision, the industry still lacks tools that provide seamless modeling. These tools should ideally support strong formal foundations for automatic analysis, code generation, and validation~\cite{Amarnath2016, Yu2015}. Additionally, early validation and verification (V\&V) of the system behavior ensure that the system design meets the requirements, reducing late discovered errors~\cite{Cederbladh2024}.

\noindent\textit{Software Design Qualities}. The distribution of dependable system components presents significant integration and interoperability challenges within the automotive industry. To address these issues, the industry is applying standardization frameworks (refer to the online supplementary material) and adopting continuous integration (CI) methods. However, the implementation of CI is complicated by the need to protect intellectual property, often restricting access to crucial information. Academically, front-loading is proposed as a solution involving modeling components and their functional and non-functional properties at earlier development stages. However, modeling the extensive number of existing "black box" components requires substantial effort~\cite{Amarnath2016}.
Many automobile variants and configurations characterize the automotive domain~\cite{Haghighatkhah2017, Knauss2016}. Approaches for variability models include feature models, decision modes, and UML-based approaches. These models can compromise up to thousands of features and result in tedious manual work~\cite{Bashroush2017, Berger2013}. Key challenges involve the evolution and visualization of models, managing dependencies, streamlining the configuration process, and ensuring traceability~\cite{Berger2013, Haghighatkhah2017, Schäfer2021}. Schäfer et al.\ highlight the need for structured approaches to managing system model variability, often less rigorously handled than source code variability, and emphasize the importance of providing decision support tools and frameworks to help practitioners select the most appropriate variability mechanisms~\cite{Schäfer2021}.

There is a high diversity of variability management tools, notations, and strategies in the industry~\cite{Berger2013}. Tools can be commercial (e.g., Microsoft Excel, Word, or Rhapsody) or based, e.g., on the Generic Eclipse Modeling Framework. The discrepancy between academia and industry seems prevalent here, with minimal collaborations. Furthermore, most studies do not thoroughly evaluate these tools' usability, integration, scalability, or performance ~\cite{Bashroush2017}. D'Ambrosio and Soremekun~\cite{Ambrosio2017} recommend a Model-Based Systems Engineering (MBSE) approach complemented by a mature change management process. This strategy ensures a thorough evaluation of dependencies and effective management of variants.\\\\

\textit{Recording Software Design}. System behavior is primarily described using SysML, UML, OCL, Simulink/MATLAB, or custom languages. Common SysML diagrams include state machine, block, sequence, and use case diagrams. However, the analysis of system behavior is often case-dependent and constrained by the narrow, domain-specific applications of these languages and formalisms, making it challenging to generalize to broader contexts. While general-purpose languages and formalisms such as SysML, MATLAB/Simulink, Modelica, and Petri nets are also utilized, discrepancies between the languages used for description and analysis rarely undergo verification for transformation correctness~\cite{Bashroush2017, Yu2015}.
Liao et al.\ propose the Vehicle Domain-Specific Language (VDSL), a low-code tool designed to enhance efficiency in vehicle functionality development. VDSL combines modeling with automated code generation to facilitate the creation of vehicle systems~\cite{Liao2024}.

\begin{keytakeaway}
\textbf{Key Takeaway:} 
The automotive software design process faces integration challenges, lack of standardization, variability management, and tool fragmentation. A seamless model-based approach with automation, early validation, CI, and a structured change management process can address these challenges.
\end{keytakeaway}

\subsubsection{Software Construction (4 Studies)}

\textit{Software Construction Fundamentals}. Today, complex systems share the same challenges: (I) the system's complexity grows faster than it can be managed, (II) the system design does not emerge from the architecture but rather from pieces of it, and (III) the knowledge and investment get lost across the project life cycle and between projects~\cite{Ambrosio2017}.  Adding to these challenges, knowledge and competence are fragmented across multiple suppliers, further complicating consistent and integrated system development~\cite{Knauss2016}. Additionally, the automotive domain faces challenges due to legacy systems' prevalence and software heterogeneity and distribution~\cite{Haghighatkhah2017}.\\\\
\textit{Practical Considerations}. In automotive system integration, ensuring performance and safety is crucial. Standardized architectures typically address early syntactical and technological challenges. However, complex issues often emerge later. Automating frequent integration processes and investing in virtualized platforms facilitates early and CI, reducing system dependencies and minimizing manual errors~\cite{Haghighatkhah2017}.
The tools used in the automotive domain show mixed support for CI. While system engineering tools provide scalability and organization-wide integration for managing current architectures, their lack of graphical modeling limits usability in some areas. Conversely, architecture model tools excel in visual modeling and adaptability but are limited in scope and integration across departments. The fragmented tool landscape, where different tools serve similar purposes in specialized contexts, creates challenges in ensuring coherence and cross-functional collaboration necessary for effective CI. This highlights the need for better integration and alignment of tools with CI practices~\cite{Knauss2016}. 
Process-related challenges such as long build times, onboarding complexity, limited test automation, and certification barriers hinder the seamless adoption of Continuous Integration/Continuous Delivery (CI/CD) in CPS development. At the same time, pipeline-related issues, including Hardware-in-the-Loop (HiL) integration, flaky tests, resource constraints, and security risks, make it difficult to maintain reliable and efficient CI/CD workflows. Addressing these challenges requires hybrid approaches that combine automation with human oversight, improved simulation techniques, and security-aware CI/CD strategies tailored to the needs of CPSs~\cite{Zampetti2023}.\\

\begin{keytakeaway}
\textbf{Key Takeaway:} Automotive software construction faces challenges in managing complexity, heterogeneity of legacy systems, distributed software components, and tool fragmentation. Automated integration, virtualized platforms, better tool alignment, and security-aware development help overcome these hurdles.
\end{keytakeaway}

\subsubsection{Software Testing (33 Studies)}

\textit{Software Testing Fundamentals}. Requirements, test management, and automation are predominantly challenging areas in automotive software testing. The recommended approaches to address these challenges include enhancing requirements and competence management, improving quality assurance, optimizing standards, optimizing test automation and tools, and incorporating agile methodologies~\cite{Haghighatkhah2017, Zimmermann2023}.
The creation of test cases is an important test issue. Test cases are also written primarily in natural language, often too lengthy or do not contain essential details~\cite{Garousi2017, Drave2018, Altinger2014, Lachmann2014}.

Natural language introduces language-based problems like translation inaccuracies and spelling errors. A significant challenge in test cases written in natural language is their phrasing. There are variants of phrasing depending on the understanding and writing style (e.g., excessive use of abbreviations, too abstract, or too much prose)~\cite{Juhke2020}. Various syntaxes, semantics, data formats, and interface descriptions hinder the exchange of test specifications amongst teams and systems, complicating collaboration and integration efforts~\cite{Haghighatkhah2017}. 
Furthermore, almost half of the specifications contain graphical elements (e.g., sketches or diagrams)~\cite{Altinger2014}. This suggests that more traditional methods like natural language and diagrams are preferred over more structured methods like pseudo-code and formal language.
The existing conformity of test case specifications (e.g., test case specification templates) fails to capture the heterogeneous nature of domains, and hard-coded test cases make reusing similar parts impossible~\cite{Juhke2020}.

According to test engineers in Drave et al.~\cite{Drave2018}, the quality focus for test cases is, e.g., functionality, robustness, or reliability. However, test case reusability, effectiveness, and maintainability were marked as less important. A similar answer is stated in the Juhnke et al. survey~\cite{Juhke2020}, where interviewees often had no idea what quality meant for their test specifications, but when quality characteristics were named, they would state comprehensibility, unambiguity, and completeness as important qualities. The less-stated characteristics are uniformity, atomicity, and suitability for the respective test platform. This mindset leads to aggravated issues, e.g., lack of documentation, updateability, knowledge management, or communication~\cite{Juhke2020, Garousi2017, Kasoju2013, Schroeder2015}. 

In driving automation systems, there appears to be a discrepancy in the general requirements. Practitioners consider performance, reliability, robustness, safety, and user comfort as important quality attributes~\cite{Habibullah2024}.

Juhnke et al.~\cite{Juhnke2022} introduced a quality model based on ISO 25010 to evaluate test case specifications, featuring seven characteristics: suitability, compatibility, usability, reliability, safety, maintainability, and portability. They also introduced a multidimensional review process with five dimensions aimed at industrial application: compliance with company guidelines, alignment with test plans, requirement correctness and coverage, test platform feasibility, and safety requirements.

Although the generation and prioritization of test cases hold significant potential, their adoption in the industry remains limited. This lack of utilization can often be attributed to requirements that are incomplete, unverifiable, unmodifiable, or unmeasurable~\cite{Atoum2021, Haghighatkhah2017, Mohd-Shafie2022, Petrenko2015}. Furthermore, Mohd-Shafie et al.~\cite{Mohd-Shafie2022} and Zampetti et al.~\cite{Zampetti2023} observe that the necessity for manual intervention and the challenges associated with implementing these techniques in large-scale systems complicate their practical application and slow down development. Petrenko et al.~\cite{Petrenko2015} address these manual intervention challenges with the Tester-in-the-Loop MBT approach, where testers interact with automated tools during crucial steps of the test case generation. This method allows testers to manually contribute or refine test case fragments, effectively combining the computational efficiency of tools with the intuition and expertise of testers.

Building on the need for more autonomous solutions, Naimi et al.~\cite{Naimi2024} propose a new approach by integrating generative AI with XML representations of use case diagrams to automate test case generation. This method promises to enhance test coverage and accuracy while reducing time-to-market. However, its effectiveness is currently limited to simpler use case diagrams, and its performance heavily depends on the AI model's training and ongoing maintenance.

Arrieta et al.~\cite{Arrieta2018} integrated formal optimization algorithms that improved fault detection, reduced execution time, and improved overall test coverage. A drawback of formal models in requirement and test cases is that stakeholders understand them worse than in semi-formal or natural language~\cite{Atoum2021, Bringmann2008}. 

\bigskip\noindent
\textit{Test Levels}. System tests in the automotive industry assess the behavior of the System under Test (SUT) to confirm it meets specified requirements. Typically, each SUT is associated with detailed requirements, specifications, and testing procedures, all documented separately from the design and requirement specifications. This separation can lead to overlooked errors or design defects~\cite{Ambrosio2017}. One significant challenge aggravating these issues is the development of accurate test beds that simulate real-world vehicle environments, which is costly and time-consuming due to vehicles' numerous ECUs. This leads to delayed testing until late in the development cycle, increasing the likelihood of discovering faults late and potentially impacting vehicle safety and compliance.
Furthermore, cross-ECU or HiL testing, crucial for validating interactions among tightly coupled ECUs, adds another layer of complexity, especially when some ECUs are still under development. The extensive data these ECUs process complicates input and output validation, further increasing safety risks with faulty inputs. These tests become more complex as they involve multiple stakeholders (testers, developers, and HiL platform specialists) who often work for different suppliers~\cite{Moukahal2020, Schroeder2015, Zampetti2023}.

Unlike functional requirements, non-functional requirements are often poorly defined. Stakeholders may be unaware of these requirements due to the need for specific domain and organizational knowledge. Gruber et al.~\cite{Gruber2017} enhance the SysML stereotype <<Requirement>> by introducing <<FunctionalRequirement>> and <<NonFunctionalRequirement>>, demonstrating the benefits of standardizing requirement specifications through clearly defined semantics.

\bigskip\noindent
\textit{Test Techniques}. The most frequent testing techniques for V\&V analysis are simulation, virtual testing, model checking, and manual inspection~\cite{Cederbladh2024}.

Knowledge-based or ML-based testing exploits or derives knowledge about the SUT. Advantages include immediate requirements feedback and shared guidelines for requirements artifacts for stakeholders. Furthermore, the automotive industry could benefit from knowledge-based testing and a controlled vocabulary to automate the identification of ambiguities and requirements errors~\cite{Atoum2021}. 
Tahir and Alexander~\cite{Tahir2020} propose a coverage-based testing framework for verifying and validating (V\&V) the safety of autonomous vehicles. This framework includes three primary coverage criteria: requirements, scenario, and situation coverage, each addressing different aspects of vehicle testing. While requirements coverage is commonly employed to ensure regulatory compliance, it cannot adequately test real-world adaptability. Scenario coverage, on the other hand, effectively tests traffic-specific events yet often fails to address edge cases comprehensively. Although less frequently utilized, situation coverage offers the most robust validation for safety-critical components within autonomous vehicles. Tahir and Alexander~\cite{Tahir2020} recommend employing a combination of these coverage criteria alongside AI-based adaptive testing strategies to enhance the robustness of testing and manage the complexities associated with autonomous driving.
Coverage-maximizing techniques are pseudorandom generation, search-based software testing, reinforcement learning, and high throughput testing. 
Pseudorandom testing continues to be the dominant approach; however, its efficiency is compromised by the lack of a targeted optimization strategy and the generation of low-value test cases.
Search-based software testing, which utilizes optimization techniques (e.g., genetic algorithms or simulated annealing), offers a more structured approach that enhances coverage efficiency. However, this method can entail significant costs.
Reinforcement learning and AI-based methods are emerging but underutilized. High throughput testing runs large-scale automated testing using cloud computing or parallel execution environments and is rarely used due to infrastructure constraints.

Wiecher et al.~\cite{Wiecher2020} propose the Scenario in the Loop (SCIL) framework, an extension of the X-in-the-Loop concept.
This approach utilizes natural language requirements specifications from Behavior-Driven Development (BDD) with formal, test-driven modeling and analysis. The SCIL framework involves a communication and documentation layer that transforms stakeholder requirements into structured scenarios using natural language and Gherkin syntax, improving collaboration between engineers and non-technical stakeholders. In the modeling layer, BDD scenarios are converted into executable models through the Scenario Modeling Language for Kotlin (SMLK), enabling early detection of requirement inconsistencies before implementation begins. SMLK-based test models are executed against real or simulated software and hardware components using industry-standard tools like Vector CANoe, MATLAB/Simulink, and winIDEA during the validation layer. This process automatically compares test results with expected behavior to confirm that system requirements are met. Despite its benefits, the SCIL framework's scalability, substantial initial setup, and the steep learning curve associated with SMLK and BDD pose challenges, particularly as automotive engineers are generally more accustomed to traditional requirement engineering methods.

\bigskip\noindent
\textit{Test-Related Measures}. Failure logic modeling (FLM) and fault injection (FI) are well-researched and found their way into several commercial software tools (e.g., FaultTree++, SafetyOffice X2, or PREEVision). However, there is no suitable set of safety mechanisms for a given system model based on the results of FLM or FI. 
Furthermore, the industry could benefit from an integrated approach of FLM and FI, especially for safety-critical functions with a high degree of automation~\cite{Amarnath2016}. 

\bigskip\noindent
\textit{Test Process}. Although a unified testing process is absent and many projects suffer from inadequate test planning, the organization has established a tester handbook that details testing processes, methods, and tools. Despite this resource, smaller teams often proceed without any formal software test plan. During the test planning phase, roles such as the customer, project manager, and test leader are actively involved. In scenarios where no test leader is available, developers themselves undertake the planning activities.
Shared roles and responsibilities between small and large teams reveal significant testing efficiency and approach differences. Smaller teams benefit from the developer performing the tester role, accelerating issue identification and resolution due to the immediate feedback loop. This reduces delays typically associated with waiting for a separate testing phase and encourages more careful coding practices. In contrast, large teams, which usually separate these roles, often do not view this as beneficial. The absence of dedicated testers in most large teams suggests potential inefficiencies and coordination challenges that could impede the testing process.

The subsequent test analysis and design phase determines the testing execution strategy. Typically, the test leader or coordinator is responsible for designing, selecting, prioritizing, and reviewing test cases. However, developers often write test cases for their code due to the shared responsibility model and the frequent unavailability of dedicated testers across projects~\cite{Kasoju2013}. Deciding which tests to run and at which development stages remains a complex issue. Testers require enhanced support in test management to effectively select appropriate tests throughout the development process, balancing the effort and early defect detection. Additional challenges include testing system of systems, calculating return on investment (ROI) for tests, developing better risk metrics, and estimating test size and effort~\cite{Garousi2017}.

Due to the flexibility of their testing activities, small teams typically do not generate extensive test reports. Conversely, large teams with structured, plan-driven approaches only produce test reports for smaller releases~\cite{Kasoju2013}. Greca et al.~\cite{Greca2023} reviewed metrics reported in the literature for measuring regression testing outcomes. Effectiveness can be measured by various metrics, including the Average Percentage of Faults Detected (APFED), testing time, accuracy, precision, recall, fault detection capability, Fault Detection Rate (FDR), and Coverage Effectiveness (CE). On the other hand, efficiency can be assessed through execution time, total or end-to-end time, memory usage, scalability, and the measurement of time or cost.

Wang et al.~\cite{Wang2022} provide practical considerations for advancing test automation maturity to ensure effectiveness, sustainability, and scalability. Key strategies include defining clear automation goals, such as improving test coverage and reducing execution time, aligning automation with business objectives, and adapting strategies to project changes. The selection of appropriate tools requires evaluating compatibility with the software under test, conducting a cost-benefit analysis of open-source versus commercial options, and assessing vendor support and tool scalability.
Effective test script development involves modular scripting for maintainability, adhering to coding standards, and implementing error-handling mechanisms. Efficient test environments leverage virtual machines and cloud testing for scalability, maintain a dedicated environment that reflects production, and ensure reliable test data. Important metrics include test execution time, pass/fail rates, defect detection ratios, script maintainability, and test automation ROI. Additionally, fostering an environment of knowledge sharing and training is crucial, involving training sessions for engineers, creating knowledge repositories, and promoting cross-team collaboration. \\\\
\textit{Software Testing in the Development Processes and the Application Domains}. There are distinct focus areas in software testing between the general industry and academia. Researchers often gravitate towards theoretically challenging issues (e.g., search-based test case design), while industry professionals prioritize improving the effectiveness and efficiency of testing practices~\cite{Garousi2017}. This difference highlights the value of collaboration between both sectors. Studies with industry practitioner co-authors often bridge the gap between theory and practice. However, feedback from practitioners of the developed tools is limited. Such feedback is crucial to evaluate long-term benefits and acceptance~\cite{Greca2023}. In addition to the focus on bridging theoretical and practical aspects of software testing, the automotive domain presents unique challenges that necessitate specific attention. Moukahal et al.~\cite{Moukahal2020} identify several distinct characteristics of vehicle software systems, such as a broad user base over a vehicle’s lifetime, numerous hardware components connected through tailored middleware, operation within critical environments, and divergent maintenance practices after sale. \\\\
\textit{Testing of and Testing Through Emerging Technologies}. Arriea et al.~\cite{Arrieta2018} even state that industrial CPSs are not testable systems, and traditional testing techniques (e.g., model-based testing) would be too expensive, time-consuming, or not feasible. Instead, simulation-based testing is anticipated to be more efficient in a systematic and automated manner. Lahami and Krichen~\cite{Lahami2021} agree with the difficulty of CPS testing, especially runtime testing of adaptable and distributed systems. In addition to the system's diverse and distributed setup and scalability issues from numerous system configurations, runtime testing is affected by the interference between testing and operational activities. They recommend employing test isolation techniques to prevent interference, optimizing test execution to conserve resources, distributing tests across different nodes to lessen the load, continuously updating test cases to keep up with system changes, and concentrating tests on critical or newly modified components to reduce the scope of testing.

Vehicle-to-vehicle, vehicle-to-infrastructure, or vehicle-to-everything (V2X) technologies represent emerging fields that facilitate interactions between independently designed automotive or similar systems. Testing V2X is crucial to prevent unpredictable behavior during these interactions. Current approaches to V2X testing incorporate MBSE, contract-based design, and the integration of formal deterministic methods with probabilistic models~\cite{Ambrosio2017}.

Developing and testing autonomous driving systems requires OEMs and suppliers to establish safety targets and comply with evolving AI and safety standards, which are complex and costly to maintain. This necessitates creating detailed safety cases to demonstrate that safety goals are met. Developers must monitor key performance metrics through tests in various driving scenarios to ensure system reliability. The unpredictable nature of ML models, which demand large volumes of high-quality data and accurate simulations, compounds these challenges. Consequently, verifying and validating autonomous systems requires rigorous statistical analysis, precise simulations, and ongoing data quality checks~\cite{Habibullah2024}.

The automotive industry faces significant challenges in testing IoT systems, primarily due to integrating many heterogeneous devices that blend diverse hardware, software, and communication standards. This integration often results in tightly coupled hardware and software components, complicating testing and maintenance. Key challenges include ensuring compatibility across multiple platforms and maintaining reliable Wi-Fi, Bluetooth, cellular networks, and GPS connectivity. These systems must manage extensive testing covering network failures and data synchronization errors.
Furthermore, automotive IoT systems must process large amounts of real-time data, necessitating robust storage and analytical frameworks. Security concerns are paramount, as these systems must implement stringent measures to safeguard against unauthorized access and ensure compliance with privacy regulations such as ISO 26262~\cite{Hagar2022, Minani2024}.
IoT testing complexities are compounded by limited access to real devices, challenges with non-standard compliant devices, difficulties reproducing bugs, fault localization, and the broad range of technologies involved. Additionally, the variable behavior of IoT devices in non-repeatable scenarios due to environmental changes or unpredictable actions can lead to inconsistent results and make issue identification challenging~\cite{Minani2024}.
Comprehensive testing strategies like Software-in-the-Loop (SIL) testing and real-world validation are crucial to manage these complexities effectively. These approaches are essential for addressing the system-of-systems complexity and ensuring that all components function correctly in diverse operational environments~\cite{Hagar2022}. \\\\
\textit{Software Testing Tools}. Testing techniques and tools are also identified challenges, especially the lack of a unified tool for all testing activities~\cite{Juhke2020, Garousi2017, Kasoju2013, Bringmann2008, Broy2010, Ambrosio2017, Knauss2016, Schroeder2015, Sztipanovits2018}. The testing toolkit comprises various tools, including test cases and data generators, test execution tools, defect detection and management tools, debugging tools, and tools for requirement traceability, configuration management, ECU modeling, and analysis. When specific tools are unavailable, practitioners often resort to customized solutions. However, smaller teams frequently rely on spreadsheets~\cite{Kasoju2013}.

The amount of tools causes interoperability, traceability, and tool support to suffer. Juhnke et al.~\cite{Juhke2020} identified the lack of overviews of test platforms and the functionalities of test platforms, with an interviewee saying \textit{"I have no idea what kind of testing levels and test platforms we have"}. Broy et al.~\cite{Broy2010} observe that engineers work with isolated tools, tailoring their processes to available resources. While the extensive array of tools in automotive development presents integration challenges, adopting a singular tool environment is equally problematic. Differences in work procedures and tool requirements between organizational sections make relying solely on one tool environment to cover all aspects of the V-model impractical. Experts recommend using tools with open APIs and well-documented file formats to ensure seamless integration and functionality.

Significant challenges arise from automation tools in testing, with a pressing need for better integration, improved metrics for automation, simplified test case automation using domain-specific languages (DSLs) or tables, and the development of fully automated test scripts that can seamlessly interact with hardware devices~\cite{Haghighatkhah2017, Altinger2014}. Testers deploy a mix of commercial, open-source, and self-developed tools that provide graphical editors, plug-in support, code generators, enhanced quality assurance capabilities, comprehensive test environments, and model creation facilities~\cite{Altinger2014}. Greca et al.~\cite{Greca2023} also highlight the limited availability of test tools after publication but acknowledge a positive trend, with more papers offering tools and replication documentation in recent years.

\begin{keytakeaway}
\textbf{Key Takeaway:}  Automotive software testing is highly complex, driven by growing system requirements and increased software dependency. The industry struggles with poorly defined test cases, often written in natural language, which leads to ambiguities, inconsistencies, and difficulties in automation. Tool fragmentation leads to interoperability and traceability between test platforms and testing levels.

The industry must move towards automated, standardized, and AI-enhanced testing processes, emphasizing formal test case specifications, reusable testing frameworks, and real-time validation techniques. A structured continuous testing and integration approach, aligned with agile and DevOps principles, is necessary to improve test efficiency, reduce costs, and enhance vehicle software reliability and safety.
\end{keytakeaway}

\subsubsection{Software Maintenance (1 Studies)}

\textit{Software Maintenance Processes}. Simulink and Stateflow have become crucial tools in automotive development, enhancing system quality and maintenance through effective clone management. Research supports the use of automated solutions to identify clones in behavioral models, with case studies confirming their effectiveness and scalability~\cite{Haghighatkhah2017}.

\begin{keytakeaway}
\textbf{Key Takeaway:} Software maintenance in the automotive industry relies on tools like Simulink and Stateflow to enhance system quality and ensure efficient long-term upkeep. Automated clone detection is crucial in managing duplicated code within behavioral models, reducing redundancy, and improving maintainability.
\end{keytakeaway}

\subsubsection{Software Engineering Management (6 Studies)}

\textit{Life Cycles}. According to the survey conducted by Altinger et al.~\cite{Altinger2014}, variations in testing methods are observed at different stages of the software development life cycle, indicating the need for adaptive testing strategies. The development within the automotive industry can be categorized into three distinct stages: Research (RE), Pre-Development (PD), and Series Development (SD). In RE, the emphasis is primarily on the simulation and the vehicle as an integrated system, considering its overall functionality and performance. In contrast, during SD, the focus shifts to individual units or components, ensuring their specific performance and reliability within the vehicle.
During the RE phase of vehicle development, a significant portion of the testing is carried out manually, accounting for approximately 65\%, in contrast to the PD phase, where manual testing comprises approximately 39\%. This figure is slightly lower in SD, about 37\%.
The discrepancy becomes more pronounced upon reflection of the testing agents. In the RE phase, it is observed that  54.29\% of the tests are conducted by the developers themselves. In contrast, in the SD phase, 55.71\% of the testing is carried out by external parties.\\\\
\textit{Software Process Assessment and Improvement}. While predominantly used in the automotive industry, the V-Model ensures rigorous verification and validation before progressing to subsequent stages and supports tailoring and outsourcing. However, its rigid nature limits flexibility and is time-consuming~\cite{Moukahal2020}. The incremental adoption of agile methods in the automotive industry promises shorter release cycles without sacrificing quality. To fully leverage these benefits, it is crucial to integrate software reuse, implement quality assurance measures, and enhance collaboration with suppliers. Such shifts necessitate transforming organizational structures and cultures. Additionally, adopting a new mindset would support advancements in online diagnostics and over-the-air updates, necessitating the development of robust infrastructure for CI and Continuous Delivery (CD) in automotive systems~\cite{Haghighatkhah2017, Knauss2016, Zampetti2023}. Small teams particularly benefit from CI and agile testing practices, which enhance communication and collaboration~\cite{Kasoju2013}.

\begin{keytakeaway}
\textbf{Key Takeaway:} Software engineering management in the automotive industry must balance structured processes like the V-Model with agile and continuous development and integration approaches to improve adaptability and efficiency. Furthermore, enhancing software reuse strategies and collaboration with suppliers improve the quality of the products. 
\end{keytakeaway}

\subsubsection{Software Engineering Models and Methods (12 Studies)}

\textit{Modeling}. A CPS model integrates many heterogeneous physical components, software, computation platforms, and network process models. The dynamic feedback loop suffers from the intricate interaction between physical and computational components. Consequently, CPSs necessitate well-defined composition semantics and a holistic modeling approach that merges various engineering disciplines, such as control engineering, software engineering, and sensor networks~\cite{Derler2012, Goswami2012, Sztipanovits2018}.
A system model captures comprehensive information and offers views tailored to stakeholders. Unlike document-based systems, the model-based approach enhances collaboration, promotes reuse, and facilitates the use of patterns. This approach reduces risks and complexity and is a reliable source of truth. Moreover, a system model can be seamlessly integrated with a requirements database or other analytical models, enabling automatic updates in response to changes in requirements~\cite{Ambrosio2017}. 
Addressing uncertainties in the measurement, expression, and propagation within CPS design models presents a significant challenge~\cite{Sztipanovits2018}. Kranabitl et al.\ introduce a framework that combines model-based system engineering with methods that define the execution of development tasks~\cite{Kranabitl2024}. This framework enhances interdisciplinary collaboration by connecting high-level system models to discipline-specific tools and methods. It ensures traceability and quality assessment throughout the development process. Sztipanovits et al.\ introduce the OpenMETA framework to improve CPSs design~\cite{Sztipanovits2018}. OpenMETA addresses challenges like heterogeneity and complexity through its Model and Tool Integration Platforms. The Model Integration Platform uses CyPhyML and a semantic backplane for precise model interactions, the Tool Integration Platform automates design and verification processes, enhancing efficiency and reducing manual errors. \\\\
\textit{Analysis of Models}. The strong interconnection of model components increases the possibility of errors. The errors can be categorized as unit, semantic, and transposition errors.
\enlargethispage{-\baselineskip}
Modeling CPSs using hybrid models, which integrate continuous-time dynamics with discrete events, presents several challenges, including solver-dependent behavior, nondeterministic behavior, and Zeno behavior. 

Solver-dependent behavior arises because continuous-time models depend on numerical solvers that approximate solutions to differential equations. These solvers dynamically adjust their time increments (step sizes) based on the system's state and numerical precision constraints. The internal mechanism for selecting these step sizes can vary between solvers, resulting in unpredictable behaviors and potentially affecting model accuracy and consistency.

Non-determinism in models can lead to multiple potential behaviors, even when the real system is deterministic. This often results from concurrent discrete events where the model's language semantics do not specify the sequence of event occurrences.

Zeno behavior occurs when infinitely many events happen in a finite amount of time, potentially freezing simulation time. This is exemplified in systems like adaptive cruise control in cars, where continuous speed adjustments can trigger endless cycles of braking and accelerating.

To address these issues, it is essential to employ numerical solvers with adaptive time-stepping and robust event detection mechanisms, redesign models to minimize dependence on frequent events and establish precise, consistent semantics for handling events that occur simultaneously. 

The challenge in modeling functionality and its implementation derives from managing time-sensitive behaviors while maintaining a clear separation between the logical operations of the system (functionality) and the details of the platform-specific execution, such as hardware constraints, network delays, and computational performance.

When CPSs are physically distributed, additional issues arise, including disparities in time measurements, concurrency, network delays, imperfect communication, and state inconsistencies~\cite{Derler2012}.

Lu et al.\ propose a toolchain that combines domain-specific modeling (DSM) and service-oriented architecture (SOA) to facilitate co-simulation and co-design for performance analysis~\cite{Lu2018}. The DSM approach uses meta-models to formalize development processes and capture system-specific characteristics. At the same time, SOA employs technical resources as services using standards such as Open Service for Lifecycle Collaboration (OSLC) and Functional Mock-up Interface (FMI).
OSLC services enhance traceability between development processes and technical resources. FMI-based co-simulation improves interoperability across multi-domain models, and the automation of repetitive tasks significantly reduces manual effort, speeding up simulation processes. However, the toolchain faces limitations due to the lack of standardization in ontology design, and it requires further testing in complex scenarios. \\\\
\enlargethispage{-\baselineskip}
\textit{Software Engineering Methods}. Model-based or model-driven development streamlines the development process and ensures the system's completeness and adherence to the expected behavior. It is the predominant test method in automotive development~\cite{Altinger2014, Bringmann2008, Drave2018, Knauss2016}. Despite the advantages of this approach, the industry frequently lacks a formal foundation for describing component structures and behaviors, which is critical for systematic system analysis. As a result, it relies more heavily on incremental development, coupled with extensive simulation and testing, to ensure the correct implementation and integration of these components~\cite{Amarnath2016}.
Broy et al.\ suggest a comprehensive modeling theory, an integrated architectural model, and a seamless model engineering environment to achieve the full promise of model-based development~\cite{Broy2010}. For the implementation of MBSE, modeling languages like Architecture Description Language (ADL), UML/SysML, and Lifecycle Modeling Language (LML) are used. The automotive industry predominantly uses SysML to implement MBSE practices~\cite{Ambrosio2017}.

\begin{keytakeaway}
\textbf{Key Takeaway:} The automotive industry relies on model-driven and CPS approaches to manage increasing software complexity. While model-based engineering enhances collaboration, reduces errors, and improves system consistency, challenges remain in semantic integration, error propagation, and interdisciplinary modeling. Future advancements require refined modeling languages, improved simulation accuracy, and better integration between formal methods and engineering tools.
\end{keytakeaway}

\subsubsection{Software Quality (1 Studies)}

\textit{Software Quality Fundamentals}. While Karhapää et al.\ did not specifically focus on the automotive industry, the parallels in organizational attributes, product structures, and processes make their insights on quality requirements (QRs) highly relevant~\cite{Karhapää2021}. In large, distributed development environments, proactive QR planning is crucial for managing complexity and ensuring team alignment. A multi-level CI framework is essential for addressing interoperability issues with legacy systems, incrementally identifying and resolving conflicts, and ensuring consistent testing and validation of integrations. Additionally, leveraging the expertise of developers enables ongoing refinement and enhancement of QRs, allowing for adaptation to emerging challenges and technological advancements.\\

\begin{keytakeaway}
\textbf{Key Takeaway:} Ensuring software quality in the automotive industry requires proactive planning, structured quality requirements, and CI frameworks to manage complexity and interoperability challenges. Developers must be actively involved in refining software quality metrics, and CI-driven validation processes should be emphasized to prevent defects and improve software robustness. The industry's ongoing shift towards software-centric vehicle architectures makes automated quality assurance, early testing, and structured QR planning essential for maintaining safety, performance, and regulatory compliance.
\end{keytakeaway}

\subsection{Requirements Analysis}
\label{sec:requirements}

The 50 primary studies extracted in Section~\ref{sec:soa} yielded 34 testing criteria.
When consolidating the coded challenges and requirements into the criteria catalog, we also consulted domain-specific documentation such as ASAM and AUTOSAR specifications to better understand current standards, terminology, and tool support. These sources were used solely to contextualize and interpret the SLR findings and were not included in the pool of 50 primary studies.

The criteria catalog was derived by open coding and consolidating the challenges and requirements extracted from the primary studies, followed by grouping them into seven requirement categories. This literature-based synthesis was complemented by feedback and experience from multiple industrial projects with OEMs and suppliers, where we iteratively refined and prioritized the criteria. These criteria were then mapped to seven thematic categories: Test Planning, Testing Tools and Infrastructure, Requirements Quality, Test Automation, Advanced Techniques, Reporting and Metrics, and Organizational Processes (see Table~\ref{tab:c_table}). We highlight the key stakeholders typically responsible for driving or benefiting from these testing criteria.

The catalog is designed as a generic framework that organizations can tailor to their specific roles, legacy toolchains, and regulatory environments, rather than as a rigid one-size-fits-all checklist. Its design is informed by both the SLR results and our experience from several cross-company projects with OEMs and suppliers, which helped us avoid criteria that are specific to a single organization.
\begingroup
\small
\rowcolors{2}{white}{gray!5}
\setlength\LTpre{0pt}\setlength\LTpost{0pt}
\begin{longtable}{
  >{\centering\arraybackslash}p{0.9cm}   
  >{\raggedright\arraybackslash}p{3.8cm} 
  >{\raggedright\arraybackslash}p{5.8cm} 
  >{\centering\arraybackslash}p{1.4cm}   
}
\caption{Detailed criteria catalog for evaluating automotive testing}
\label{tab:c_table}\\
\toprule
\textbf{ID} & \textbf{Criteria} & \textbf{Ref.} & \textbf{Prio.} \\
\midrule
\endfirsthead
\multicolumn{4}{@{}l}{\small\slshape Table \thetable\ (continued)}\\[-4pt]
\toprule
\textbf{ID} & \textbf{Criteria} & \textbf{Ref.} & \textbf{Prio.} \\
\midrule
\endhead
\bottomrule
\endlastfoot

\addlinespace[3pt]
\multicolumn{4}{l}{\textbf{Test Planning and Scope Management}} \\[3pt]
C$_1$ & Model-based approach  
      &~\cite{Yu2015,Haghighatkhah2017,Ambrosio2017,Kranabitl2024,Altinger2014,Bringmann2008,Drave2018,Knauss2016,Broy2010} 
      & P$_1$ \\
C$_2$ & Variants and configuration management
      & ~\cite{Atoum2021,Haghighatkhah2017,Knauss2016,Bashroush2017,Berger2013,Schäfer2021,Ambrosio2017} 
      & P$_1$ \\
C$_3$ & Requirements alignment
      & ~\cite{Habibullah2024,Karhapää2021,Masmoudi2022,Moukahal2020,Schroeder2015,Tukur2021, Zahid2022,Zimmermann2023,Juhke2020,Ambrosio2017}
      & P$_1$ \\

\addlinespace[3pt]
\multicolumn{4}{l}{\textbf{Testing Tools and Infrastructure Requirements}} \\[3pt]
C$_4$ & Unified and integrated test environment 
      &~\cite{Juhke2020,Garousi2017,Kasoju2013,Bringmann2008,Broy2010,Ambrosio2017,Knauss2016,Schroeder2015,Sztipanovits2018} 
      & P$_1$ \\
C$_5$ & Interoperability and open standards 
    & ~\cite{Altinger2014,Amarnath2016,Ambrosio2017,Bringmann2008,Broy2010,Feiler2012,Garousi2017,Gruber2017,Habibullah2024,Haghighatkhah2017,Juhke2020,Karhapää2021,Kasoju2013,Knauss2016,Masmoudi2022,Moukahal2020,Schroeder2015,Sztipanovits2018,Tukur2021,Yu2015,Zahid2022,Zimmermann2023} 
      & P$_1$ \\
C$_6$ & Scalability and performance
      & ~\cite{Ambrosio2017,Bashroush2017,Knauss2016,Mohd-Shafie2022,Zampetti2023,Tahir2020,Wiecher2020,Greca2023,Wang2022,Lahami2021,Haghighatkhah2017,Moukahal2020,Schroeder2015}
      & P$_1$ \\
C$_7$ & User-centric tooling
      & ~\cite{Liao2024,Bashroush2017,Yu2015,Juhke2020,bruel2020,Knauss2016,Zampetti2023,Haghighatkhah2017,Altinger2014,Kasoju2013,Garousi2017,Wang2022}
      & P$_2$ \\
C$_8$ & End-to-end traceability
      & ~\cite{Berger2013,bruel2020,Garousi2017,Haghighatkhah2017,Juhke2020,Kasoju2013,Kranabitl2024,Lu2018,Norheim2024,Schäfer2021,Tukur2021,Zahid2022,Zimmermann2023}
      & P$_1$ \\

\addlinespace[3pt]
\multicolumn{4}{l}{\textbf{Requirements and Specification Quality}} \\[3pt]
C$_{9}$ & Clarity and standardization of specifications
         & ~\cite{Lachmann2014,Zhao2021,Drave2018,Gruber2017,Habibullah2024,Tukur2021,Yu2015,Zahid2022,Altinger2014} 
         & P$_1$ \\
C$_{10}$ & Multi‑format specification support
         & ~\cite{Haghighatkhah2017,Altinger2014,Juhke2020,Atoum2021,Bringmann2008,Wiecher2020} 
         & P$_3$ \\
C$_{11}$ & NLP support for specifications
         & ~\cite{Altinger2014,Atoum2021,bruel2020,Drave2018,Garousi2017,Gruber2017,Habibullah2024,Haghighatkhah2017,Juhke2020,Kasoju2013,Lachmann2014,Masmoudi2022,Norheim2024,Tukur2021,Wiecher2020,Yu2015,Zahid2022,Zhao2021}
         & P$_2$ \\
C$_{12}$ & Template and Modularization 
         & ~\cite{Juhke2020,Garousi2017,Kasoju2013,Schroeder2015,Juhnke2022,Drave2018} 
         & P$_3$ \\
C$_{13}$ & Completeness and Consistency 
         & ~\cite{Ambrosio2017,Atoum2021,Gruber2017,Habibullah2024,Haghighatkhah2017,Juhke2020,Masmoudi2022,Mohd-Shafie2022,Petrenko2015,Tukur2021} 
         & P$_1$ \\
C$_{14}$ & Verifiability and Testability 
         & ~\cite{Masmoudi2022,bruel2020,Atoum2021,Haghighatkhah2017,Mohd-Shafie2022,Petrenko2015}
         & P$_1$ \\
C$_{15}$ & Visual-element integration 
         & ~\cite{Altinger2014,Knauss2016,Bashroush2017,Yu2015,Haghighatkhah2017}   
         & P$_2$ \\
C$_{16}$ & Legacy‑system handling
         & ~\cite{Habibullah2024,Karhapää2021,Masmoudi2022,Moukahal2020,Schroeder2015,Tukur2021,Zahid2022,Zimmermann2023,Haghighatkhah2017} 
         & P$_2$ \\

\addlinespace[3pt]
\multicolumn{4}{l}{\textbf{Test Automation and Execution}} \\[3pt]
C$_{17}$ & Automation framework capabilities
         &~\cite{Altinger2014,Arrieta2018,Atoum2021,Drave2018,Garousi2017,Haghighatkhah2017,Juhke2020,Kasoju2013,Mohd-Shafie2022,Naimi2024,Petrenko2015,Schroeder2015,Tahir2020,Wang2022,Zampetti2023,Zhao2021,Zimmermann2023} 
         & P$_1$ \\
C$_{18}$ & Automated test case generation 
         & ~\cite{Petrenko2015,Naimi2024,Haghighatkhah2017,Zimmermann2023,Tahir2020} 
         & P$_1$ \\
C$_{19}$ & High-throughput and parallel execution  
         & ~\cite{Hagar2022,Zampetti2023,Cederbladh2024,Tahir2020,Atoum2021,Wiecher2020,Wang2022,Lahami2021,Habibullah2024,Lu2018,Amarnath2016}
       & P$_1$ \\
C$_{20}$ & CI/CD integration 
         & ~\cite{Zampetti2023,Amarnath2016,Haghighatkhah2017,Knauss2016,Lahami2021,Kasoju2013,Karhapää2021} 
         & P$_1$ \\
C$_{21}$ & Robust scripting and error handling
         & ~\cite{Derler2012,Wang2022,Haghighatkhah2017,Altinger2014,Lu2018,Liao2024}
         & P$_1$ \\

\addlinespace[3pt]
\multicolumn{4}{l}{\textbf{Advanced Techniques and Technologies}} \\[3pt]
C$_{22}$ &Integration of advanced test techniques
         & ~\cite{Altinger2014,Amarnath2016,Ambrosio2017,Arrieta2018,Atoum2021,Bringmann2008,Broy2010,Derler2012,Drave2018,Feiler2012,Habibullah2024,Knauss2016,Masmoudi2022,Naimi2024,Norheim2024,Yu2015,Zhao2021}
         & P$_2$ \\

C$_{23}$ & AI-driven optimization 
         & ~\cite{Naimi2024,Zhao2021,Norheim2024,Atoum2021,Haghighatkhah2017,Mohd-Shafie2022,Petrenko2015} 
         & P$_2$ \\
C$_{24}$ & Simulation environments 
         & ~\cite{Ambrosio2017,Hagar2022,Moukahal2020,Schroeder2015,Zampetti2023,Tahir2020}
         & P$_1$ \\
C$_{25}$ & Research alignment 
         & ~\cite{Garousi2017,Greca2023,Bashroush2017,Atoum2021,Haghighatkhah2017,Mohd-Shafie2022,Petrenko2015} 
         & P$_2$ \\

\addlinespace[3pt]
\multicolumn{4}{l}{\textbf{Reporting, Metrics, and Continuous Improvement}} \\[3pt]
C$_{26}$ & Reporting dashboards 
         & ~\cite{Wang2022,Kasoju2013,Greca2023}
         & P$_3$ \\
C$_{27}$ & Automated analysis and visualization 
         & ~\cite{Wiecher2020,Garousi2017,Greca2023,Wang2022}
         & P$_2$ \\
C$_{28}$ & Feedback loops 
         & ~\cite{Ambrosio2017,Knauss2016,Haghighatkhah2017,Zimmermann2023,Juhnke2022}
         & P$_2$ \\

\addlinespace[3pt]
\multicolumn{4}{l}{\textbf{Organizational Processes, Collaboration, and Agile Practices}} \\[3pt]
C$_{29}$ & Stakeholder alignment and communication 
         & ~\cite{Juhke2020,Ambrosio2017,Tukur2021,Zahid2022,Masmoudi2022,Atoum2021,Bringmann2008,Gruber2017,Wiecher2020, Garousi2017,Habibullah2024}
         & P$_1$ \\
C$_{30}$ & Agile and iterative processes 
         & ~\cite{Haghighatkhah2017,Zimmermann2023,Knauss2016,Zampetti2023,Kasoju2013,Moukahal2020}
         & P$_3$ \\
C$_{31}$ & Cross-functional collaboration 
         & ~\cite{Knauss2016,Wang2022,Altinger2014,Tukur2021,Juhke2020,Ambrosio2017,Moukahal2020,Schroeder2015,Zampetti2023,Kasoju2013,Gruber2017}
         & P$_2$ \\
C$_{32}$ & Defined roles and responsibilities
         & ~\cite{Derler2012,Goswami2012,Sztipanovits2018,Ambrosio2017,Tukur2021,Zahid2022,Kasoju2013}
         & P$_2$ \\
C$_{33}$ & Continuous process improvement
         & ~\cite{Tukur2021,Zahid2022,Atoum2021,Masmoudi2022,Yu2015,Kasoju2013,Derler2012,Goswami2012,Sztipanovits2018}
         & P$_2$ \\
C$_{34}$ & Knowledge sharing 
         & ~\cite{Amarnath2016,Ambrosio2017,Knauss2016,Juhke2020,Garousi2017,Kasoju2013,Schroeder2015,Wang2022} 
         & P$_2$ \\

\end{longtable}
\endgroup

\textbf{Test Planning and Scope Management} consists of three criteria primarily influenced by OEMs and Tier-1 suppliers due to the necessity of managing complex, variant-rich product lines. This category emphasizes the importance of a model-based approach to plan tests, where scenarios and system behaviors are integrated into system models, supporting early validation and dynamic adaptability (C$_{1}$). It also highlights stringent handling of automotive-specific variants and configurations, essential for preventing redundant or missed tests and ensuring comprehensive coverage (C$_{2}$). Moreover, maintaining continuous alignment with evolving project requirements through tight integration with requirement management systems prevents outdated test information and enhances test effectiveness (C$_{3}$).

\textbf{Testing Tools and Infrastructure} includes five criteria essential for OEM toolchains and platform providers. Specifically, it emphasizes the need for unified and integrated environments encompassing test design, automation, execution, analysis, and reporting to reduce fragmentation and streamline processes (C$_{4}$). Interoperability and compliance with open standards are critical, requiring tools to offer open APIs, standardized file formats, and extensibility mechanisms to ensure seamless integration with legacy systems, simulation platforms, and CI/CD pipelines (C$_{5}$). The category also stresses scalable performance characteristics suitable for large-scale, distributed testing scenarios, including multi-ECU environments and simulation setups like HiL and SiL (C$_{6}$). User-centric tool design, featuring intuitive interfaces, customizable dashboards, and low-code or domain-specific languages, enhances accessibility and collaboration (C$_{7}$). Finally, comprehensive end-to-end traceability among requirements, test cases, and outcomes is essential, supported by detailed metrics-driven reporting and visualization tools to promote continuous improvement (C$_{8}$).

\textbf{Requirements and Specification Quality}, comprising eight criteria, is relevant to all stakeholders in the development process, ranging from systems engineers to software providers. Requirements should adhere to standards such as ISO/IEC/IEEE 29148, ensuring clarity, consistency, and avoiding ambiguities (C$_{9}$). Supporting multiple specification representations, e.g., natural language, semi-formal notations, formal models, domain-specific languages, and graphical diagrams, accommodates diverse development stages and stakeholder expertise (C$_{10}$). Integrating NLP-based tools enhances the detection of ambiguities and inconsistencies (C$_{11}$). Reusable specification templates facilitate atomicity, maintainability, and uniformity, streamlining maintenance and reducing redundancy (C$_{12}$). Ensuring completeness and consistency across documentation and communications is critical for comprehensive coverage and predictability (C$_{13}$). Requirements must be verifiable, measurable, and traceable to test cases for adequate validation and impact analysis (C$_{14}$). Visual elements like state machines and sequence diagrams should integrate formally with textual specifications to support diverse communication preferences (C$_{15}$). Lastly, addressing interoperability with legacy systems is essential to prevent integration challenges and missed test scenarios (C$_{16}$).
In practice, NLP-based support for requirements engineering in the automotive domain is still emerging and most often reported in prototypes and pilot studies rather than in large-scale production use; accordingly, we treat such techniques as promising directions rather than mature, widely deployed solutions.

\textbf{Test Automation and Execution} encompasses five criteria valued by OEMs and suppliers specializing in test automation. This category advocates the adoption of robust automation frameworks capable of supporting complex automotive testing environments (C$_{17}$). Automated test case generation should integrate formal and AI-driven techniques such as search-based methods, reinforcement learning, and generative AI, tailored specifically for automotive systems to maximize test coverage and reduce manual effort (C$_{18}$). High-throughput and parallel execution capabilities, especially in cloud-based and simulation environments like SiL and HiL, are vital for accurately replicating real-world automotive scenarios (C$_{19}$). Seamless integration with CI/CD pipelines ensures rapid feedback and early defect detection, essential for maintaining vehicle safety and minimizing time-to-market delays (C$_{20}$). Lastly, reliable scripting and robust error management mechanisms using modular, maintainable scripts in domain-specific languages ensure long-term reliability and ease of maintenance as systems evolve (C$_{21}$).
However, both the primary studies and our insights from industrial collaborations indicate that AI-driven test generation and optimization are currently adopted mainly in research prototypes and limited industrial pilots; portfolio-level scalability in production projects remains an open issue.

\textbf{Advanced Techniques and Technologies}, with four criteria, frequently emerges from the efforts of research institutions and forward-thinking suppliers. This includes applying formal methods to improve accuracy and reliability (C$_{22}$). AI-based optimization strategies are crucial for enhancing test case prioritization, maximizing coverage, and reducing redundant or low-value tests, especially in managing complex automotive systems and the uncertainties associated with ML components in autonomous applications (C$_{23}$). Realistic and comprehensive simulation environments are essential for accurately testing multi-ECU systems, vehicle-to-everything interactions, and cyber-physical scenarios, reflecting real-world variability and safety-critical conditions (C$_{24}$). Strategic alignment between academia and industry through joint research initiatives, technology transfers, pilot projects, and workshops ensures theoretically robust and practically applicable methodologies, accelerating their adoption in the automotive testing landscape (C$_{25}$).
Across our corpus, these advanced techniques are predominantly reported in experimental studies, tool demonstrators, or early-stage industrial pilots; mature, large-scale industrial uptake is comparatively rare, which is why we classify them as exploratory or emerging directions in the criteria catalog.

\textbf{Reporting and Metrics}, consisting of three criteria, is driven primarily by quality managers and project leaders. This category prioritizes the development of rich reporting dashboards that display critical metrics such as defect detection rates (including APFED and FDR), test execution times, coverage effectiveness, and the ROI from test automation, enabling rapid, data-driven decision-making (C$_{26}$). Automated analysis and visualization tools should assess test results systematically and generate visual reports tracking progress over time to support ongoing improvements and justify investments in testing (C$_{27}$). Explicit mechanisms for integrating feedback from testers, developers, and stakeholders ensure continuous test specifications and tool performance refinement, creating an iterative enhancement cycle (C$_{28}$).

\textbf{Organizational Processes and Agile Practices} involve six criteria jointly managed by project management teams, agile coaches, and various suppliers. It advocates for broad stakeholder alignment and effective communication (C$_{29}$) and adopting agile methodologies such as Scrum and Kanban to integrate testing activities early in development. Agile and iterative workflows support shorter release cycles and rapid feedback, facilitating early defect detection and flexible test planning responsive to evolving requirements (C$_{30}$). Cross-functional team collaboration among testers, developers, domain experts, and suppliers ensures comprehensive and nuanced test case design and execution (C$_{31}$). Clearly defined roles and responsibilities prevent ambiguity and promote accountability, aligning testing efforts with project objectives (C$_{32}$). Continuous process improvement practices supported by metrics-driven retrospectives foster a learning culture that enhances test quality and efficiency (C$_{33}$). Structured knowledge-sharing mechanisms, including centralized repositories and enhanced communication channels such as daily stand-ups and retrospectives, minimize duplication and ensure vital insights are accessible to all stakeholders (C$_{34}$).

Finally, the qualitative priority labels (P$_{1}$ - P$_{3}$) were assigned based on a combination of literature citations, insights from practitioner interviews, and the authors’ cumulative automotive-testing experience.

The next subsection presents the numerical results and discusses how the composite ranking supports the focus of our second literature review.

\section{Specification Techniques and Tools}
\label{sec:specs_tools}

Test case specification techniques determine \textit{what} is tested, while testing tools govern \textit{how} efficiently those specifications are executed and maintained across platforms. Evidence from our SLR (Section~\ref{sec:slr}) shows that reported barriers cluster in these two layers and recur most frequently in the citation review. Focusing this second review on specification techniques and supporting toolchains aligns the scope with the most frequently reported problems in academia and practice and with the priorities established by the catalog.

This section provides a systematized overview of specification techniques and tools. Section~\ref{sec:method_languages_tools} describes our identification method; Sections~\ref{sec:results_languages_tools} and~\ref{sec:results_testing_tools} present the resulting sets of techniques and tools. Complete references to tools, frameworks, and companies are provided in the supplementary material for completeness.

\subsection{Method}
\label{sec:method_languages_tools}
For this comprehensive review, we adopt the PRISMA framework~\cite{Liberati2009} to make the mixed evidence base explicit, since the corpus includes non-database sources (e.g., vendor documentation, release notes, open-source repositories, technical blogs) alongside scholarly databases.

We established eligibility criteria to guide the review and ensure the inclusion of relevant and recent sources:

\begin{itemize}
  \item[]\textbf{IC-1 Relevance of Content}: The title, abstract, or keywords must include at least one of the following terms: 'Test Case,' 'Specification,' 'Language,' 'Automotive,' 'Testing Tools,' or 'System Testing.'
  \item[]\textbf{IC-2 Type of Publication}: The work must be a journal articles, conference papers, or workshop papers.
  \item[]\textbf{IC-3 Language}: The publication must be in English.
  \item[]\textbf{IC-4 Publication Date}: The work must have been published between 2021 and 2025, as tools and technologies in computer science evolve rapidly and are often deprecated within a short time frame.
  \item[]\textbf{IC-5 Citation Impact}: To prioritize influential studies and to keep screening tractable, we required a minimum of 10 citations for academic publications. Items from 2024–2025 were exempt to mitigate recency bias. This criterion does not apply to non-database sources (e.g., vendor documentation, standards, release notes, open-source repositories, technical blogs).
\end{itemize}

We also defined the following exclusion criteria:
\begin{itemize}
  \item[]\textbf{EC-1 Availability}: The publication, specification, or tool is unavailable.
  \item[]\textbf{EC-2 Applicability}: The publication is theoretical without practical application.
  \item[]\textbf{EC-3 Review Status}: The publication is a preprint or non-peer-reviewed paper, including early results or preliminary studies.
\end{itemize}
Typical inclusions for the PRISMA-guided review were, for instance, papers that introduced a specification language and illustrated it with an automotive case study, or tool descriptions that reported automotive deployments or examples. We also included non-database sources, such as vendor documentation, when they provided sufficient technical details on test workflows and automotive usage. In contrast, we excluded generic test management or CI tools that lacked any automotive evidence, as well as purely promotional web pages without technical content. As in the SLR, borderline cases were discussed between the reviewers, with a preference for inclusion at the first pass and re-evaluation during full-text analysis.

Guided by IC-1, we defined two query templates that combine three concept blocks: (i) domain (automotive), (ii) target (specification techniques or testing tools), and (iii) synonyms/aliases with wildcards and proximity where supported. We instantiated these templates per database and documented all run dates and filters in Table~\ref{tab:prisma_sting_database_1} and Table~\ref{tab:prisma_sting_database_2}.

\begin{description} 
    \item[SS1 Specification techniques.] (("test case" NEAR/3 (specification OR language OR design OR description)) OR "specification language" OR "domain-specific language" OR DSL OR DSML OR "model-based test*" OR MBT) AND (automotive OR vehicle* OR ADAS OR "in-vehicle" OR "in vehicle" OR ECU OR SDV OR "software-defined vehicle" OR "software defined vehicle") 
    \item[SS2 Testing tools.] (("testing tool*" OR "test platform" OR "test management" OR "test automation" OR "continuous integration" OR CI OR "continuous delivery" OR CD) AND (automotive OR vehicle* OR ADAS OR "in-vehicle" OR "in vehicle" OR ECU OR SDV OR "software-defined vehicle" OR "software defined vehicle")) 
\end{description}

We conducted searches across multiple scholarly databases, IEEE Xplore DL, Google Scholar, DBLP, Scopus, and ACM DL, to ensure comprehensive coverage:
\begingroup
\footnotesize
\rowcolors{2}{white}{gray!5}
\setlength\LTpre{0pt}
\setlength\LTpost{0pt}
\begin{longtable}{
@{}
>{\bfseries}p{0.14\textwidth}
p{0.80\textwidth}
@{}
}
\caption{PRISMA Database-specific queries for SQ1} 
\label{tab:prisma_sting_database_1} \\
\toprule
\textbf{Database} & \textbf{Search Query} \\
\midrule
\endfirsthead

\multicolumn{2}{@{}l}{\small\slshape Table \thetable\ (continued)} \\[-3pt]
\toprule
\textbf{Database} & \textbf{Search Query} \\
\midrule
\endhead

\bottomrule
\endlastfoot

 IEEE Xplore DL&(("All Metadata":"test case specification" OR "All Metadata":"specification language" OR "All Metadata":"domain-specific language" OR "All Metadata":"model-based test*" OR "All Metadata":MBT) AND ("All Metadata":automotive OR "All Metadata":vehicle OR "All Metadata":"in-vehicle" OR "All Metadata":ADAS OR "All Metadata":ECU OR "All Metadata":SDV OR "All Metadata":"software-defined vehicle" OR "All Metadata":"software defined vehicle"))
Refinements: Publication Years 2021–2025; Content Type = Journals, Conferences. Language = English.
 \\

Google Scholar & Ran as 4 separate queries to avoid query truncation:
\begin{itemize}
    \item "test case specification" automotive OR vehicle OR ADAS OR "in-vehicle" OR ECU OR SDV OR "software-defined vehicle" OR "software defined vehicle"
    \item "specification language" automotive OR vehicle OR ADAS OR "in-vehicle" OR ECU OR SDV OR "software-defined vehicle" OR "software defined vehicle"
    \item "domain-specific language" test case automotive OR vehicle OR ADAS OR "in-vehicle" OR ECU OR SDV OR "software-defined vehicle" OR "software defined vehicle"
    \item "model-based testing" "test case" automotive OR vehicle OR ADAS OR "in-vehicle" OR ECU OR SDV OR "software-defined vehicle" OR "software defined vehicle"
\end{itemize}
 \\

DBLP &    \begin{itemize}
       \item "test case specification" automotive
       \item "test case specification" "software-defined vehicle"
       \item "test case specification" "software defined vehicle"
       \item "specification language" automotive
       \item "specification language" "software-defined vehicle"
       \item "specification language" "software defined vehicle"
       \item "domain-specific language" test case
       \item "model-based testing" "test case" automotive
   \end{itemize}
 \\

Scopus &TITLE-ABS-KEY(
 (("test case" W/3 (specification OR language OR design OR description))
  OR "specification language" OR "domain-specific language" OR DSL OR DSML
  OR "model-based test*" OR MBT)
 AND (automotive OR vehicle* OR ADAS OR "in-vehicle" OR "in vehicle" OR ECU OR SDV OR "software-defined vehicle" OR "software defined vehicle")
)
AND PUBYEAR > 2020 AND PUBYEAR < 2026 AND LANGUAGE(English)
 \\
ACM DL&("test case specification" OR "specification language" OR "domain-specific language" OR DSL OR DSML OR "model-based testing" OR MBT) AND (automotive OR vehicle OR ADAS OR "in-vehicle" OR ECU OR SDV OR "software-defined vehicle" OR "software defined vehicle"))\\
\end{longtable}
\endgroup

\begingroup
\footnotesize
\rowcolors{2}{white}{gray!5}
\setlength\LTpre{0pt}
\setlength\LTpost{0pt}
\begin{longtable}{
@{}
>{\bfseries}p{0.14\textwidth}
p{0.80\textwidth}
@{}
}
\caption{PRISMA Database-specific queries for SQ2} 
\label{tab:prisma_sting_database_2} \\
\toprule
\textbf{Database} & \textbf{Search Query} \\
\midrule
\endfirsthead

\multicolumn{2}{@{}l}{\small\slshape Table \thetable\ (continued)} \\[-3pt]
\toprule
\textbf{Database} & \textbf{Search Query} \\
\midrule
\endhead

\bottomrule
\endlastfoot

 IEEE Xplore DL&(("All Metadata":"testing tools" OR "All Metadata":"test platform" OR "All Metadata":"test management" OR "All Metadata":"test automation" OR "All Metadata":"continuous integration" OR "All Metadata":CI OR "All Metadata":"continuous delivery" OR "All Metadata":CD) AND ("All Metadata":automotive OR "All Metadata":vehicle OR "All Metadata":"in-vehicle" OR "All Metadata":ADAS OR "All Metadata":ECU OR "All Metadata":SDV OR "All Metadata":"software-defined vehicle" OR "All Metadata":"software defined vehicle")) 
Refinements: Publication Years 2021–2025; Content Type = Journals, Conferences; Language = English. 
 \\

Google Scholar  &Ran as 3 separate queries to avoid query truncation:
\begin{itemize}
    \item "testing tools" automotive OR vehicle OR ADAS OR "in-vehicle" OR ECU OR SDV OR "software-defined vehicle" OR "software defined vehicle"
    \item "test platform" automotive OR vehicle OR ADAS OR "in-vehicle" OR ECU OR SDV OR "software-defined vehicle" OR "software defined vehicle"
    \item "test automation" automotive OR vehicle OR ADAS OR "in-vehicle" OR ECU OR SDV OR "software-defined vehicle" OR "software defined vehicle"
\end{itemize} 
 \\

 DBLP&    \begin{itemize}
       \item "testing tools" automotive
       \item "testing tools" "software-defined-vehicle"
       \item "testing tools" "software defined vehicle"
       \item "test platform" automotive
       \item "test platform" "software-defined-vehicle"
       \item "test platform" "software defined vehicle"       
       \item "test automation" automotive
       \item "test automation" "software-defined-vehicle"
       \item "test automation" "software defined vehicle"
   \end{itemize}
 \\

Scopus& TITLE-ABS-KEY( ( "testing tool*" OR "test platform" OR "test management" OR "test automation" OR "continuous integration" OR CI OR "continuous delivery" OR CD ) AND (automotive OR vehicle* OR ADAS OR "in-vehicle" OR "in vehicle" OR ECU OR SDV OR "software-defined vehicle" OR "software defined vehicle") ) AND PUBYEAR > 2020 AND PUBYEAR < 2026 AND LANGUAGE(English)
 \\
ACM DL& ("testing tools" OR "test platform" OR "test management" OR "test automation" OR "continuous integration" OR CI OR "continuous delivery" OR CD) AND (automotive OR vehicle OR ADAS OR "in-vehicle" OR ECU OR SDV OR "software-defined vehicle" OR "software defined vehicle"))
 \\
\end{longtable}
\endgroup

We executed structured Google queries using the SS1/SS2 concept terms, scanning the first 50 results per query (sorted by relevance). Eligible items were vendor documentation, standards pages, release notes, and open-source repositories with substantive technical content. We recorded query strings and run dates and retained stable URLs.

Screening and data extraction for SS1 and SS2 followed the same collaborative procedure as in the SLR. Two authors first applied the inclusion and exclusion criteria to a pilot set of records to calibrate decisions and refine the written guidelines. The remaining items were then divided between the authors for screening, with borderline cases discussed and resolved by consensus. Information about specification techniques and testing tools was extracted into a shared spreadsheet and reviewed by a second author for accuracy. As in the SLR, we did not compute a formal inter-rater statistic; instead, we relied on iterative calibration and consensus.

We merged exports from each source and removed duplicates by DOI when available and otherwise by normalized title and year. We archived the raw exports and a deduplicated master file. Searches were run between the first of  October 2024 and the 31st of March 2025; citation counts for IC-5 were snapshotted on the first of February 2025 from Google Scholar. We archived raw exports and a deduplicated master file to ensure reproducibility.

\begin{figure}
    \centering
    \includegraphics[width=0.9\linewidth]{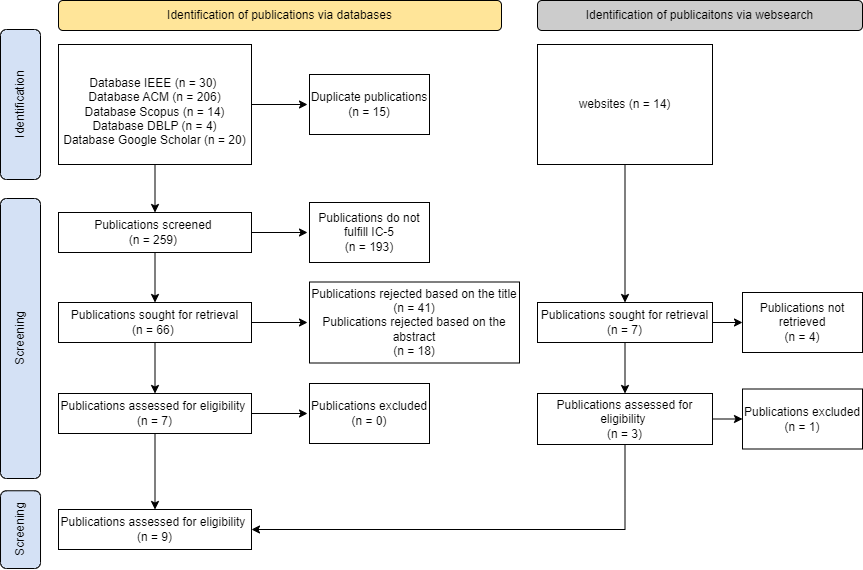}
    \caption{Screening process and selection outcomes using search string SS1}
    \label{fig:Prisma1}
\end{figure}

The search with \textit{SS1} identified 274 records; see Figure~\ref{fig:Prisma1}. The PRISMA flow diagram reports separate counts for database and gray sources
and quantifies the effect of IC-5 on the academic subset.
After removing 15 duplicates and excluding 193 papers based on IC-5, 66 records remained. Screening by title reduced this to 41, followed by abstract screening to 18, and finally, seven papers were selected for full-text analysis. 
A complementary Google search identified 14 websites, of which six were examined in detail. Four were excluded, leaving two suitable sources. Combined with the academic papers, this resulted in nine relevant sources.

An analysis of these nine sources identified six specification options for test cases: (1) Cucumber's Gherkin, (2) ETSI's TTCN-3, and (3) OEM's UML/SysML were each mentioned in three separate sources ~\cite{dosSantos, JuhnkeTichy2019, Micallef2015, Juhnke2022, bruel2020, Elekes2023}; specifications in the form of a (4) table, (5) constraint logic programming (CLP) languages, and (6) ASAM test specifications were each mentioned once respectively~\cite{dosSantos, Pretschner2005, ASAM}.
%

\begin{figure}
    \centering
    \includegraphics[width=0.9\linewidth]{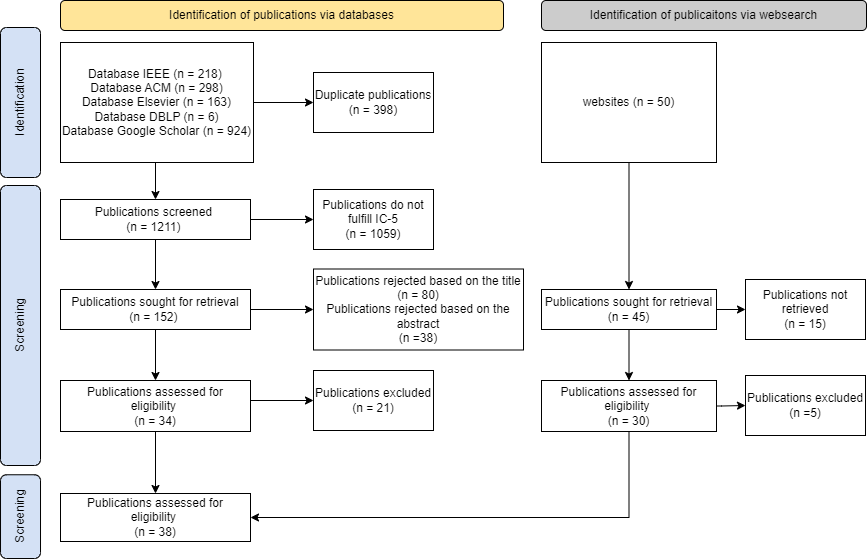}
    \caption{Screening process and selection outcomes using search string SS2}
    \label{fig:Prisma2}
\end{figure}

%
Screening \textit{SS2} identified 1,609 records in five databases; see Figure~\ref{fig:Prisma2}.  After removing 398 duplicates and excluding 1,059 based on IC-5, 152 records remained. Title and abstract screening reduced this to 34 papers for full-text review, with 13 ultimately included.
A complementary Google search identified 50 websites, 45 of which were preliminarily assessed. After a detailed review, 25 met the inclusion criteria. In total, 38 sources were selected for this study.

Analyzing these 38 sources yielded five recurring tool categories. 

\textbf{Test Platform} tools emerged as the predominant category, referencing seven different tools, covering aspects such as requirements engineering and data management~\cite{Rotaru2024, Ebert2022, jenkins, Oka2019, CANoe, Cucumber, ttworkbench, Mohd-Shafie2022, ude}. \textbf{X-in-the-Loop (XiL)} tools and frameworks were analyzed in four instances, focusing on various simulation environments, including SiL, HiL, and Vehicle-in-the-Loop (ViL)~\cite{vTESTstudio, Ebert2022, CANoe, dSPACE, Zsolt2021, Mihalic2022, Abboush2022, alma}. Six \textbf{Automation} tools were examined or mentioned in the results~\cite{robotframework, Peldszus2023, Minani2024, Mohd-Shafie2022, selenium, Garousi2021, Ebert2022, keysight, softing}.

Furthermore, seven \textbf{Modeling} tools supporting Model-Based Testing (MBT) were mentioned in the results~\cite{mathworks, Kim2015, Szalay2021, Abboush2022, Saini2022, preevision, stateflow, Sinha2021, Zafar2021, openmeta}. Lastly, four \textbf{Quality Management} tools were identified~\cite{alm, Ebert2022, Atoum2021}.
Multifunctional tools were categorized based on their primary usage. For example, if a test platform includes quality control functions but is primarily used for developing and deploying tests, it is categorized as 'Test Platform' rather than 'Quality Management.'

\subsection{Specification Techniques and Tools Research Results}
\label{sec:results_languages_tools}

\subsubsection{Test Case Specification Techniques}

\paragraph{Gherkin} Gherkin is a domain-specific language for writing acceptance tests in a structured "Given–When–Then" format. Its natural language syntax supports clear communication of requirements by non-technical stakeholders while enabling automated execution. dos Santos and Vilain highlight that Gherkin reduces documentation effort without sacrificing clarity, making it effective in agile and behavior-driven development~\cite{dosSantos}. Its readability fosters collaboration across roles and helps standardize test case creation, promoting consistency and shared understanding throughout development~\cite{Micallef2015}.

\paragraph{TTCN-3} Testing and Test Control Notation version 3 (TTCN-3) is a mature, standardized testing language used in domains like automotive and distributed systems. It is a platform-independent bridge between modeling and execution, supporting the automated validation of system conformance and performance through its formal syntax and semantics. Juhnke et al. show that integrating TTCN-3 with quality models like ISO 9126 improves test specification clarity and consistency~\cite{Juhnke2022}. Its structured nature makes it well-suited for large-scale, safety-critical environments requiring concurrent test execution and early error detection~\cite{JuhnkeTichy2019}.

\paragraph{SysML/UML} UML and its systems-oriented extension, SysML, use graphical notations to describe system behavior, requirements, and test scenarios in model-based testing. Clear semantics, especially in UML state machines, are crucial for generating understandable and executable test cases~\cite{Elekes2023}. SysML extends this with requirements diagrams that improve traceability between system elements and support functional and non-functional specifications. Their visual representation helps clarify complex interactions, making them particularly useful in safety-critical domains~\cite{bruel2020}.

\paragraph{Table} Tabular test specifications like Fit Tables organize test scenarios in a clear row-and-column format, detailing inputs, actions, and expected outcomes~\cite{fittable}. They are easy to learn and apply, offering a consistent framework for documenting acceptance tests~\cite{dosSantos}. This format reduces ambiguity, supports stakeholder comprehension, and facilitates automation by easing the conversion of test cases into executable scripts.

\paragraph{CLP Language} CLP combines logic programming and constraint-solving to automate test case generation from behavioral models~\cite{Pretschner2005}. Models are translated into logical conditions, enabling symbolic execution over sets of possible values. This reduces manual effort, improves error detection, and integrates constraint reasoning into model-based testing. Though not a complete replacement for manual testing, CLP is a strong complement, particularly when evolving requirements demand efficient test regeneration.

\paragraph{ASAM} The Association for Standardization of Automation and Measuring Systems (ASAM) Test Specification standard defines a structured framework for automotive test case development and execution, emphasizing consistency, traceability, and tool interoperability. It is especially vital for safety-critical systems like Advanced Driver Assistance Systems (ADAS) and Automated Driving (AD). The ASAM Automotive Test Exchange Format (ATX) format standardizes test data exchange, supporting the reuse of test cases and enabling manual and automated procedures for efficient test automation~\cite{ASAM}.

\subsubsection{Testing Tools}
\label{sec:results_testing_tools}
In this section, we present various tools identified through the PRISMA method. XiL tools facilitate early fault detection and component validation across integration levels (Table~\ref{tab:xil_description}), while automation frameworks enhance test execution efficiency and coverage (Table~\ref{tab:auto_description}). Modeling tools aid in system-level analysis and design quality (Table~\ref{tab:model_description}), and data management solutions ensure effective handling of test data for CI (Table~\ref{tab:data_description}). Additionally, quality management software provides traceability and oversight of requirements, defects, and testing activities (Table~\ref{tab:quali_description}).

\begingroup
\footnotesize
\rowcolors{2}{white}{gray!5}  
\setlength\LTpre{0pt}\setlength\LTpost{0pt}

\begin{table}[htbp]
  \centering
  \begin{tabularx}{\textwidth}{@{}               
      >{\raggedright\arraybackslash}p{3.5cm}     
      >{\raggedright\arraybackslash}X            
      >{\centering\arraybackslash}p{2 cm}     
    @{}}
    \toprule
    \textbf{Tool Name} & \textbf{Description} & \textbf{Refs.} \\
    \midrule

    [HiL] vTeststudio& Tool for test development and automation across Model-in-the-Loop (MiL) to HiL in automotive embedded systems.  &~\cite{Ebert2022}  \\
\addlinespace[2pt]
    [SiL + HiL] CANoe& Widely used for simulation, testing, and analysis of vehicle communication networks and real-time ECU testing. & ~\cite{CANoe} \\
\addlinespace[2pt]
    [XiL] dSPACE ASM& Model library for simulating and validating automotive systems at MiL, SiL, Process-in-the-Loop (PiL), and HiL levels. & ~\cite{Zsolt2021,Mihalic2022,dSPACE,Abboush2022,Ebert2022} \\
\addlinespace[2pt]
    [XiL] VeriStand&Platform for real-time testing, integration, and validation, offering broad XiL support. & ~\cite{Mihalic2022,Ebert2022} \\

    \bottomrule
  \end{tabularx}
  \caption{X-in-the-Loop tool descriptions}
  \label{tab:xil_description}
\end{table}
\endgroup

\begingroup
\footnotesize
\rowcolors{2}{white}{gray!5}  
\setlength\LTpre{0pt}\setlength\LTpost{0pt}

\begin{table}[htbp]
  \centering
  \begin{tabularx}{\textwidth}{@{}               
      >{\raggedright\arraybackslash}p{3.5cm}     
      >{\raggedright\arraybackslash}X            
      >{\centering\arraybackslash}p{2 cm}     
    @{}}
    \toprule
    \textbf{Tool Name} & \textbf{Description} & \textbf{Refs.} \\
    \midrule
    Robot Framework
    & Open-source framework for acceptance testing (ATDD) and robotic process automation.
    & ~\cite{Peldszus2023,Minani2024,Mohd-Shafie2022} \\
\addlinespace[2pt]
    Selenium
    & Web automation tool for testing web applications across browsers.
    & ~\cite{Minani2024,Garousi2021,Mohd-Shafie2022,Ebert2022} \\
\addlinespace[2pt]
    VectorCAST 
    & Tool for automating unit and integration testing with code coverage for embedded and safety-critical systems.
    & ~\cite{Minani2024,Ebert2022} \\
\addlinespace[2pt]
    Apache JMeter 
    & Tool for load, performance, and functional testing of web services. 
    & ~\cite{Minani2024} \\
\addlinespace[2pt]
    Citrus Framework
    & Framework for integration testing of messaging-based systems; supports BDD tools like Cucumber.
    & ~\cite{Ebert2022} \\
\addlinespace[2pt]
    Keysight's Eggplant
    & AI-based UX testing platform supporting CI/CD integration. 
    & ~\cite{keysight} \\

    \bottomrule
  \end{tabularx}
  \caption{Automation Framework / Software Tool Descriptions}
  \label{tab:auto_description}
\end{table}
\endgroup

\begingroup
\footnotesize
\rowcolors{2}{white}{gray!5}  
\setlength\LTpre{0pt}\setlength\LTpost{0pt}

\begin{table}[htbp]
  \centering
  \begin{tabularx}{\textwidth}{@{}               
      >{\raggedright\arraybackslash}p{3.5cm}     
      >{\raggedright\arraybackslash}X            
      >{\centering\arraybackslash}p{2 cm}     
    @{}}
    \toprule
    \textbf{Tool Name} & \textbf{Description} & \textbf{Refs.} \\
    \midrule    
    MATLAB Simulink& Block diagram environment for model-based design and simulation, offering a graphical editor, libraries, and solvers for dynamic systems. & ~\cite{Kim2015,Szalay2021,Abboush2022,Saini2022} \\
\addlinespace[2pt]
    PREEvision& Tool for model-based E/E system engineering in automotive, supporting design, modeling, and management of architectures and networks.&~\cite{preevisionpdf}  \\
\addlinespace[2pt]
    MATLAB Stateflow& Simulates decision logic using diagrams and tables; helpful in modeling reactive systems and detecting system faults.&~\cite{Saini2022}  \\
\addlinespace[2pt]
    IBM RSA& The Rational Software Architect Designer tool for visual modeling and architecture design, supporting UML and model-driven development for complex systems.&~\cite{Sinha2021}  \\
\addlinespace[2pt]
    Eclipse Papyrus& Open-source tool for UML and SysML modeling, tailored for complex systems engineering on the Eclipse platform.&~\cite{Sinha2021}  \\
\addlinespace[2pt]
    GraphWalker & Tool for model-based testing using finite-state machine graphs; enables automated, structured test generation and execution. &  ~\cite{Zafar2021}\\
\addlinespace[2pt]
    OpenMETA&Open-source CPSs design toolset supporting integrated modeling, analysis, and simulation across engineering disciplines. & ~\cite{openmeta} \\
    \bottomrule
  \end{tabularx}
  \caption{Modeling Tool Descriptions}
  \label{tab:model_description}
\end{table}
\endgroup

\begingroup
\footnotesize
\rowcolors{2}{white}{gray!5}  
\setlength\LTpre{0pt}\setlength\LTpost{0pt}

\begin{table}[htbp]
  \centering
  \begin{tabularx}{\textwidth}{@{}               
      >{\raggedright\arraybackslash}p{3.5cm}     
      >{\raggedright\arraybackslash}X            
      >{\centering\arraybackslash}p{2 cm}     
    @{}}
    \toprule
    \textbf{Tool Name} & \textbf{Description} & \textbf{Refs.} \\
    \midrule 
    IBM Doors Next Generation & Comprehensive tool for managing, tracing, and analyzing complex requirements throughout the product life cycle. & ~\cite{Rotaru2024,Ebert2022}  \\
\addlinespace[2pt]
    jenkins&Widely used open-source automation server supporting CI/CD, automating build, test, and deployment processes. & ~\cite{Oka2019,jenkins} \\
\addlinespace[2pt]
    Cucumber&Open-source BDD tool using Gherkin syntax to write human-readable test scenarios for improved team collaboration.&~\cite{Cucumber} \\
\addlinespace[2pt]
    FitNesse& Wiki-based framework for writing and executing automated acceptance tests in a collaborative, human-readable format.& ~\cite{Ebert2022} \\
\addlinespace[2pt]
    UDE&Software tool for debugging, testing, and automating embedded systems and microcontroller development.&~\cite{ude}\\
\addlinespace[2pt]
    codeBeamer& Application Lifecycle Management (ALM) platform integrates requirements, development, testing, and compliance into one system. & ~\cite{Oka2019} \\
\addlinespace[2pt]
    CANalyzer&Tool for simulating, analyzing, and testing automotive communication networks, especially CAN-based systems. & ~\cite{Ebert2022} \\

    \bottomrule
  \end{tabularx}
  \caption{Data Management / Test Platform Tool Descriptions}
  \label{tab:data_description}
\end{table}
\endgroup

\begingroup
\footnotesize
\rowcolors{2}{white}{gray!5}  
\setlength\LTpre{0pt}\setlength\LTpost{0pt}

\begin{table}[htbp]
  \centering
  \begin{tabularx}{\textwidth}{@{}               
      >{\raggedright\arraybackslash}p{3.5cm}     
      >{\raggedright\arraybackslash}X            
      >{\centering\arraybackslash}p{2 cm}     
    @{}}
    \toprule
    \textbf{Tool Name} & \textbf{Description} & \textbf{Refs.} \\
    \midrule 
    HP ALM& Platform manages the full software life cycle, focusing on quality, governance, and traceability across requirements, defects, and tests. & ~\cite{alm} \\
\addlinespace[2pt]
    SpiraTest& The test management platform offers real-time bug tracking, exploratory testing, and reporting, with a high setup effort and moderate automation.& ~\cite{Ebert2022} \\
\addlinespace[2pt]
   MaramaAIC& Requirements quality tool using semi-formal models to detect inconsistencies between requirements. & ~\cite{Atoum2021} \\
\addlinespace[2pt]
    TestMEReq& Tool for checking requirement correctness, completeness, and consistency.&~\cite{Atoum2021}  \\

\bottomrule
  \end{tabularx}
  \caption{Quality Management Software Tool Descriptions}
  \label{tab:quali_description}
\end{table}
\endgroup

\subsection{Specification and Tool Analysis}
\label{sec:evaluation}
We use the criteria catalog introduced in Table~\ref{tab:c_table} to assess the selected test case specifications techniques systematically.  
Each criterion is rated Low (\raisebox{-0.5ex}{\emptycirc}),
Medium (\raisebox{-0.5ex}{\halfcirc}) or High (\raisebox{-0.5ex}{\fullcirc}).
A dash (–) means that the literature and vendor documentation contained
no applicable evidence for that combination, and therefore, no rating could be assigned.  
We convert the symbols into numeric
scores (High = 2, Medium = 1, Low = 0, – = n/a) for a concise ranking and sum them across all
criteria. Every table's bottom row $\Sigma$ shows the resulting total per specification or tool.

\subsubsection{Specifications}
The \textit{Requirements and Specification Quality} category stands out as particularly significant, warranting a detailed analysis, see Table~\ref{tab:cc_specificaiton}.

\begingroup
\footnotesize
\setlength\LTpre{0pt}\setlength\LTpost{0pt}

\begin{table}[htbp]
\renewcommand\arraystretch{1.05}
  \centering
  \begin{tabularx}{\textwidth}{@{}
      >{\centering\arraybackslash}p{1.2cm}   
      >{\raggedright\arraybackslash}X          
      *{6}{>{\centering\arraybackslash}p{0.75cm}}   
      @{}}
    \toprule
\rot{\textbf{Prioritized ID}} & \textbf{Criteria} &
\rot{\textbf{Gherkin}} &
\rot{\textbf{TTCN-3}} &
\rot{\textbf{SysML/UML}} &
\rot{\textbf{Table}} &
\rot{\textbf{CLP}} &
\rot{\textbf{ASAM}}\\
    \midrule
        \multicolumn{8}{@{}l}{\textbf{Testing Tools and Infrastructure Requirements}}\\[2pt]
                 C$_{8}$P$_{1}$&End-to-end traceability &  \fullcirc&    \fullcirc&    \fullcirc&       \fullcirc&     \halfcirc&   \fullcirc\\
    \multicolumn{8}{@{}l}{\textbf{Requirements and Specification Quality}}\\[2pt]
        C$_{9}$P$_{1}$ & Clarity and standardization of specifications&  \fullcirc&    \fullcirc&    \halfcirc&   \fullcirc&     \halfcirc&   \fullcirc \\
        C$_{10}$P$_{3}$ & Multi-format specification support &  \halfcirc&  \halfcirc&  \fullcirc&       \halfcirc&   \halfcirc&   \fullcirc \\
        C$_{11}$P$_{2}$ & NLP support for specifications &  \fullcirc&    \emptycirc&     \halfcirc&     \halfcirc&  -&      - \\
        C$_{12}$P$_{3}$ & Template and Modularization  &  \fullcirc&    \fullcirc&    \fullcirc&       \fullcirc&     \halfcirc& \fullcirc \\
        C$_{13}$P$_{1}$& Completeness and Consistency  &  \halfcirc&  \fullcirc&  \halfcirc&   \halfcirc&   \fullcirc&     \fullcirc \\
        C$_{14}$P$_{1}$&Verifiability and Testability &  \fullcirc&    \fullcirc&    \halfcirc&   \fullcirc&     \fullcirc&     \fullcirc \\
        C$_{15}$P$_{2}$& Visual-element integration &  \emptycirc&     \emptycirc&     \fullcirc&       \halfcirc&   \emptycirc&     \halfcirc \\
        C$_{16}$P$_{2}$&Legacy-system handling &  \fullcirc&    \halfcirc&  \fullcirc&       \fullcirc&     \halfcirc&    \halfcirc \\
\midrule[0.3pt]                   
\textbf{$\Sigma$} & \multicolumn{1}{r}{\textbf{Score}} &
\textbf{14} & \textbf{12} & \textbf{14} & \textbf{14} &
\textbf{9} & \textbf{14} \\[-0.1em]
\bottomrule

  \end{tabularx}
  \caption{Coverage of prioritized criteria from the \emph{Testing Tools and Infrastructure} and \emph{Requirements and Specification Quality} categories by six test‑case specification techniques. Ratings are High (\fullcirc), Medium (\halfcirc), Low (\emptycirc), or not applicable (–). The bottom row $\Sigma$ lists the aggregated scores per technique, where High = 2, Medium = 1, Low =0.}

  \label{tab:cc_specificaiton}
\end{table}
\endgroup
C$_{8}$ is largely rated High, as most specifications offer built-in or integrable support for requirements traceability. CLP is the exception, rated Medium due to its limited explicit support for traceability.

C$_{9}$ is rated Medium to High across all specifications. Gherkin’s structured natural language aids readability and stakeholder comprehension. TTCN-3 offers formal precision but may be less accessible to non-experts. UML/SysML diagrams are intuitive yet often require textual support. Table-based formats are clear and widely understandable. CLP is highly precise but abstract for non-technical users. ASAM’s structured format supports clear communication and aligns with industry standards.

C$_{10}$ is generally rated Medium. Most specifications (Gherkin, TTCN-3, Table-based, and CLP) use a single primary format. UML/SysML combines visual and textual elements, while ASAM supports multiple automotive-relevant formats, earning higher ratings. C$_{11}$ varies widely. Gherkin scores High, UML/SysML and Table formats reach Medium, while TTCN‑3 shows Low support, and CLP and ASAM have no reported evidence. 

C$_{12}$ is rated High for Gherkin, TTCN‑3, UML/SysML, Table formats, and ASAM. CLP achieves a Medium rating because its template mechanisms are powerful but less user-friendly. C$_{13}$ are split into Medium and High ratings, as specifications support but do not ensure these qualities. Gherkin and Tables require careful detailing; UML/SysML may need supplementary text; TTCN-3 and CLP reach High offering stronger formal support; ASAM explicitly targets comprehensive coverage in automotive contexts.

C$_{14}$ is rated High for all specifications, except UML/SysML, which receives a Medium rating due to its reliance on specific tools for effective test generation. 

For C$_{15}$, UML/SysML scores High for its inherently visual modeling approach. ASAM and Table-based specifications include some visual structuring and are rated Medium. Gherkin, TTCN-3, and CLP, being primarily textual or formal, are rated Low. 

C$_{16}$ must be evaluated in the context of individual systems; hence, the ratings provided represent estimations. UML/SysML, Tables, and Gherkin support straightforward legacy integration, while TTCN-3 and ASAM require more training and effort for adoption in legacy systems.

Gherkin, SysML/UML, Table‑based formats, and ASAM tie for first place with 14 points, TTCN‑3 follows with 12 points, and CLP attains 9 points. While the aggregate helps rank the languages at a glance, individual criteria remain decisive for project‑specific choices.

\begin{keytakeaway}
\textbf{Key Takeaway.} 
No single specification dominates across all 10 assessed criteria.  
Four techniques (Gherkin, SysML/UML, Table formats, ASAM) share the top Σ score of 14.  
ASAM and TTCN‑3 excel in standardization, modularization, and testability, which is attractive for safety‑critical domains but implies steeper onboarding.  
Gherkin supports clarity and stakeholder communication, making it a good fit for agile teams.  
Table formats remain the most accessible choice and integrate well with legacy artefacts.  
SysML/UML offers the strongest visual expressiveness and traceability, but requires modeling expertise and tooling.  
CLP achieves the highest formal completeness yet scores lowest overall because of limited human readability.  
Projects combining a human‑readable notation (Gherkin or Tables) with a formally rigorous one (TTCN‑3 or CLP) can cover more criteria without excessive complexity.
\end{keytakeaway}

\subsubsection{Tools}
\subsubsubsection{X-in-the-Loop}
\begingroup
\footnotesize  
\setlength\LTpre{0pt}\setlength\LTpost{0pt}

\begin{table}[htbp]
\renewcommand\arraystretch{1.05}
  \centering
  \begin{tabularx}{\textwidth}{@{}
      >{\centering\arraybackslash}p{1.2cm}   
      >{\raggedright\arraybackslash}X          
      *{4}{>{\centering\arraybackslash}p{0.9cm}}   
      @{}}
    \toprule
\rot{\textbf{Prioritized ID}} & \textbf{Criteria} &
\rot{\textbf{vTest\-studio}} &
\rot{\textbf{CANoe}} &
\rot{\textbf{dSPACE ASM}} &
\rot{\textbf{VeriStand}}\\
\midrule
\multicolumn{6}{@{}l}{\textbf{Test Planning and Scope Management}}\\[1pt]
C$_{1}$P$_{1}$  & Model-based approach                           & \halfcirc  & \emptycirc & \fullcirc & \fullcirc \\
C$_{2}$P$_{1}$ & Variants and configuration management           & \halfcirc  & \halfcirc  & \fullcirc & \fullcirc \\
\addlinespace[1pt]
\multicolumn{6}{@{}l}{\textbf{Testing Tools and Infrastructure Requirements}}\\[1pt]
C$_{5}$P$_{1}$  & Interoperability and open standards             & \halfcirc  & \halfcirc  & \halfcirc & \halfcirc \\
C$_{6}$P$_{1}$  & Scalability and performance                     & \halfcirc  & \fullcirc  & \fullcirc & \fullcirc \\
C$_{8}$P$_{1}$  & End-to-end traceability    & \fullcirc  & \fullcirc  & \halfcirc & \fullcirc \\

\addlinespace[1pt]
\multicolumn{6}{@{}l}{\textbf{Test Automation and Execution}}\\[1pt]
C$_{17}$P$_{1}$ & Automation framework capabilities                        & \fullcirc  & \fullcirc  & \halfcirc & \fullcirc \\
C$_{18}$P$_{1}$& Automated test case generation                & \fullcirc  & \fullcirc  & \halfcirc & \fullcirc \\
C$_{19}$P$_{1}$& High-throughput and parallel execution         & \halfcirc  & \halfcirc  & \fullcirc & \fullcirc \\
\addlinespace[1pt]
\multicolumn{6}{@{}l}{\textbf{Advanced Techniques and Technologies}}\\[1pt]
C$_{22}$P$_{2}$& Integration of advanced test techniques        & \fullcirc  & \halfcirc  & \fullcirc & \fullcirc \\

C$_{24}$P$_{1}$& Simulation of environments         & \emptycirc & \halfcirc  & \fullcirc & \fullcirc \\

\midrule[0.3pt]                   
\textbf{$\Sigma$} & \multicolumn{1}{r}{\textbf{Score}} &
\textbf{13} & \textbf{13} & \textbf{16} & \textbf{19} \\[-0.1em]
\bottomrule

  \end{tabularx}
  \caption{Coverage of prioritized criteria from four categories of the Criteria Catalog by four X-in-the-Loop tools. Ratings are High (\fullcirc), Medium (\halfcirc), Low (\emptycirc), or not applicable (–). The bottom row $\Sigma$ lists the aggregated scores per technique, where High = 2, Medium = 1, Low =0.}
  \label{tab:cc_xil_tools}
\end{table}
\endgroup

vTESTstudio excels in automation, test case design, and traceability but depends heavily on CANoe for full functionality and lacks real-world simulation capabilities.
CANoe is strong in ECU diagnostics, network analysis, and real-time testing but is less advanced in model-based techniques than ASM and VeriStand.
dSPACE ASM offers comprehensive modeling and detailed simulation environments, making it particularly suitable for real-world simulations and research.
VeriStand offers acceptable interoperability, automation, model integration, and real-time execution, though it is slightly less specialized in vehicle dynamics than dSPACE ASM, as outlined in Table~\ref{tab:cc_xil_tools}.

The comparative evaluation reveals that VeriStand achieved the highest score (19), followed by ASM (16), while vTESTstudio and CANoe received equal scores (13). This quantitative assessment confirms that VeriStand provides the most balanced overall coverage, whereas ASM demonstrates superior capabilities in modeling and simulation. The parity between vTESTstudio and CANoe highlights their complementary strengths: vTESTstudio is particularly strong in automation functionalities, while CANoe excels in real-time network analysis.
\begin{keytakeaway}
\textbf{Key Takeaway.}
With a score of 19/20 (ten criteria, High = 2), VeriStand offers the broadest coverage by combining model‑based execution, rich automation, and high‑throughput real‑time operation.  
dSPACE ASM follows at 16/20 and is preferred when physics‑based vehicle or environment simulation is critical.  
vTESTstudio and CANoe reach 13/20: vTESTstudio excels in automation and test‑case design but relies on CANoe for low‑level bus interaction. At the same time, CANoe provides first‑rate diagnostics and timing accuracy but lags in model‑centric workflows.  
No single tool scores High on every criterion; coupling a general‑purpose platform such as VeriStand with a specialist tool such as CANoe can yield fuller coverage without adding unnecessary overlap.
\end{keytakeaway}

\subsubsubsection{Automation Tools}
\begingroup
\footnotesize  
\setlength\LTpre{0pt}\setlength\LTpost{0pt}

\begin{table}[htbp]
\renewcommand\arraystretch{1.05}
  \centering
  \begin{tabularx}{\textwidth}{@{}
      >{\centering\arraybackslash}p{1.2cm}   
      >{\raggedright\arraybackslash}X          
      *{6}{>{\centering\arraybackslash}p{0.8cm}}@{}}
\toprule
\rot{\textbf{Prioritized ID}} & \rot{\textbf{Criteria}} & \rot{\textbf{Robot Framework}} & \rot{\textbf{Selenium}} & \rot{\textbf{VectorCAST}}& \rot{\textbf{Apache Jmeter}}& \rot{\textbf{Citrus Framework}}& \rot{\textbf{Eggplant}}\\
\midrule
    \multicolumn{8}{l}{\textbf{Testing Tools and Infrastructure Requirements}}\\[1pt]
        C$_{4}$P$_{1}$& Unified and integrated environment&\fullcirc&    \emptycirc&       \fullcirc&       \halfcirc&     \halfcirc&  \halfcirc \\ 
        C$_{5}$P$_{1}$&Interoperability and open standards& \fullcirc&    \fullcirc&      \halfcirc&     \fullcirc&       \fullcirc&  \halfcirc \\ 
        C$_{6}$P$_{1}$&Scalability and performance& \halfcirc&  \halfcirc&    \fullcirc&       \fullcirc&       \halfcirc&  \fullcirc \\ 
        C$_{7}$P$_{2}$& User-centric tooling&\halfcirc&  \emptycirc&       \halfcirc&     \emptycirc&        \emptycirc& \fullcirc  \\ 
        C$_{8}$P$_{1}$&End-to-end traceability& \halfcirc&  \emptycirc&       \fullcirc&       \halfcirc&     \halfcirc&  \fullcirc \\ 

\addlinespace[1pt]
    \multicolumn{8}{l}{\textbf{Test Automation and Execution}} \\[1pt]
    C$_{17}$P$_{1}$&Automation framework capabilities& \fullcirc&    \fullcirc&      \fullcirc&       \fullcirc&       \fullcirc&  \fullcirc \\ 
    C$_{18}$P$_{1}$&Automated test case generation& \halfcirc& \emptycirc&      \fullcirc&        \emptycirc&\halfcirc & \halfcirc  \\ 
    C$_{19}$P$_{1}$&High-throughput and parallel execution& \halfcirc& \halfcirc&   \halfcirc&      \fullcirc&\halfcirc & \fullcirc  \\ 
    C$_{20}$P$_{1}$& CI/CD integration&\fullcirc&   \fullcirc&     \halfcirc&      \fullcirc& \fullcirc&  \fullcirc \\ 
    C$_{23}$P$_{1}$& Robust scripting and error handling&\halfcirc& \halfcirc&   \halfcirc&      \halfcirc& \fullcirc&  \halfcirc \\ 
\midrule[0.3pt]                   
\textbf{$\Sigma$} & \multicolumn{1}{r}{\textbf{Score}} &
\textbf{14} & \textbf{9} & \textbf{15} & \textbf{13} &
\textbf{13} & \textbf{16} \\[-0.1em]
\bottomrule

  \end{tabularx}
  \caption{Coverage of prioritized criteria from the \emph{Testing Tools and Infrastructure} and \emph{Test Automation and Execution} categories by six automation tools. Ratings are High (\fullcirc), Medium (\halfcirc), Low (\emptycirc), or not applicable (–). The bottom row $\Sigma$ lists the aggregated scores per technique, where High = 2, Medium = 1, Low =0.}

  \label{tab:cc_autotools}
\end{table}
\endgroup

Robot Framework supports Windows, macOS, and Linux, allowing easy integration into existing environments. It leverages many libraries and supports data-driven testing, making it well-suited for automation and CI/CD workflows.
Selenium enables parallel test execution across multiple machines and browsers and supports several programming languages (e.g., Python, Java, C\#). In the automotive context, its use is generally limited to UI testing for infotainment systems and lacks traceability features.
VectorCAST is widely adopted in the automotive sector, supporting multiple testing levels. It offers code coverage analysis and data flow checks for safety compliance and integrates well with CI/CD pipelines, making it ideal for safety-critical software.
JMeter specializes in load and performance testing, including response time and throughput. It supports protocols (e.g., HTTP, SOAP, REST) and CI tool integration, ensuring interoperability and alignment with open standards. However, it is not designed for UI testing or complex test logic.
Citrus Framework simulates client-server interactions for message-based systems, enabling robust end-to-end integration testing. It also supports mocking of external systems (e.g., SOAP services), making it effective for testing microservices, APIs, and middleware.
Eggplant is a proprietary AI-driven test automation platform. It generates intelligent test paths based on user behavior and system logic and supports full-stack testing across UI, APIs, databases, and networks. It integrates with standard CI/CD tools and includes performance monitoring capabilities, see Table~\ref{tab:cc_autotools}.

Among the evaluated tools, Eggplant demonstrates the highest criterion coverage (score: 16), followed by VectorCAST (15), Robot Framework (14), Citrus and JMeter (both 13), and Selenium (9).
Eggplant thus offers the broadest criterion coverage, while VectorCAST leads among tools certified for safety‑critical code. Selenium’s lower score reflects its narrow focus on UI regression.
\begin{keytakeaway}
\textbf{Key Takeaway.}
With 16/20 points, Eggplant provides the widest feature coverage, combining AI‑guided test generation, cross‑stack automation, and strong scalability.
VectorCAST follows at 15/20 and remains the reference choice for unit‑ and integration‑level testing in safety‑critical projects, thanks to built‑in coverage and data‑flow analysis.
Robot Framework reaches 14/20; its rich library ecosystem and scriptable keyword style suit continuous‑integration pipelines, but traceability and high‑throughput execution are only moderate.
Citrus and JMeter share 13/20: Citrus specialises in message‑driven end‑to‑end tests with good CI/CD hooks, while JMeter excels in load and performance automation through a plug‑in architecture.
Selenium scores 9/20, reflecting its focus on browser‑based UI regression rather than full‑stack automation.
No single tool covers all ten criteria; pairing a general‑purpose platform such as Eggplant or Robot Framework with a specialist tool (for example, JMeter for load or VectorCAST for safety compliance) can widen coverage without redundant overlap.
\end{keytakeaway}

\subsubsubsection{Modeling Tools}
\begingroup
\footnotesize
\setlength\LTpre{0pt}\setlength\LTpost{0pt}

\begin{table}[!htbp]
\renewcommand\arraystretch{1.05}
  \centering
  \begin{tabularx}{\textwidth}{@{}
      >{\centering\arraybackslash}p{1.2cm}         
      >{\raggedright\arraybackslash}X                
      *{7}{>{\centering\arraybackslash}p{0.70cm}}    
  @{}}                                               
    \toprule
 \rot{\textbf{Prioritized ID}} & \textbf{Criteria} &
      \rot{\textbf{Simulink}} &
      \rot{\textbf{PREE\-vision}} &
      \rot{\textbf{Stateflow}} &
      \rot{\textbf{IBM RSA}} &
      \rot{\textbf{Papyrus}} &
      \rot{\textbf{Graph\-Walker}} &
      \rot{\textbf{OpenMETA}} \\
    \midrule
    \multicolumn{9}{@{}l}{\textbf{Test Planning and Scope Management}}\\[1pt]
    C$_{1}$P$_{1}$ & Model-based approach                         & \fullcirc & \fullcirc  & \fullcirc & \fullcirc & \fullcirc & \fullcirc & \fullcirc \\
    \addlinespace[1pt]
    \multicolumn{9}{@{}l}{\textbf{Testing Tools and Infrastructure Requirements}}\\[1pt]
    C$_{4}$P$_{1}$  & Unified and integrated test environment            & \fullcirc & \fullcirc  & \halfcirc & \fullcirc & \halfcirc & \halfcirc & \fullcirc \\
    C$_{5}$P$_{1}$  & Interoperability and open standards           & \halfcirc & \fullcirc  & \halfcirc & \fullcirc & \fullcirc & \halfcirc & \fullcirc \\
    C$_{6}$P$_{1}$ & Scalability and performance                   & \fullcirc & \fullcirc  & \fullcirc & \halfcirc & \halfcirc & \halfcirc & \fullcirc \\
    C$_{7}$P$_{2}$ & User-centric tooling   & \fullcirc & \halfcirc  & \halfcirc & \halfcirc & \fullcirc & \halfcirc & \fullcirc \\
    C$_{8}$P$_{1}$  & End-to-end traceability    & \fullcirc & \fullcirc  & \halfcirc & \fullcirc & \halfcirc & \halfcirc & \halfcirc \\

    \addlinespace[1pt]
    \multicolumn{9}{@{}l}{\textbf{Test Automation and Execution}}\\[1pt]
     C$_{17}$P$_{1}$  & Automation framework capabilities                      & \fullcirc & \fullcirc  & \fullcirc & \halfcirc & \halfcirc & \fullcirc & \fullcirc \\
    C$_{18}$P$_{1}$& Automated test case generation  & \fullcirc & \halfcirc  & \halfcirc & \halfcirc & \emptycirc & \fullcirc & \halfcirc \\
    C$_{19}$P$_{1}$& High-throughput and parallel execution       & \fullcirc & \halfcirc  & \halfcirc & \emptycirc & \emptycirc & \halfcirc & \halfcirc \\
     C$_{20}$P$_{1}$& CI/CD integration                           & \fullcirc & \halfcirc  & \halfcirc & \halfcirc & \halfcirc & \fullcirc & \halfcirc \\
    C$_{21}$P$_{1}$& Robust scripting and error handling          & \fullcirc & \halfcirc  & \halfcirc & \halfcirc & \emptycirc & \fullcirc & \halfcirc \\
    \addlinespace[1pt]
    \multicolumn{9}{@{}l}{\textbf{Advanced Techniques and Technologies}}\\[1pt]
     C$_{22}$P$_{2}$& Integration of advanced test techniques     & \fullcirc & \fullcirc  & \halfcirc & \halfcirc & \emptycirc & \fullcirc & \halfcirc \\
    C$_{23}$P$_{1}$& AI-driven optimization                      & \halfcirc & \emptycirc & \emptycirc & \emptycirc & \emptycirc & \emptycirc & \halfcirc \\
    C$_{24}$P$_{1}$& Simulation of real-world environments       & \fullcirc & \fullcirc  & \halfcirc & \halfcirc & \fullcirc  & \halfcirc & \fullcirc \\
     C$_{25}$P$_{2}$& Research alignment     & \fullcirc & \halfcirc  & \halfcirc & \halfcirc & \fullcirc  & \halfcirc & \fullcirc \\
\midrule[0.3pt]                   
\textbf{$\Sigma$} & \multicolumn{1}{r}{\textbf{Score}} &
\textbf{28} & \textbf{22} & \textbf{17} & \textbf{17} &
\textbf{15} & \textbf{20}& \textbf{23} \\[-0.1em]
\bottomrule

  \end{tabularx}
  \caption{Coverage of prioritized criteria from four categories by seven modeling tools. Ratings are High (\fullcirc), Medium (\halfcirc), Low (\emptycirc), or not applicable (–). The bottom row $\Sigma$ lists the aggregated scores per technique, where High = 2, Medium = 1, Low =0.}

  \label{tab:cc_modeltools}
\end{table}
\endgroup
OpenMETA excels in integration, customization, and interoperability, making it well-suited for multidisciplinary design analysis and optimization (MDAO). Its strong automation framework, though, auto‑generation is only Medium.
Simulink offers strong automation, advanced testing (e.g., SIL, PIL), and robust integration with CI/CD workflows. Its scripting support and real-world simulation capabilities are ideal for traceable, complex system testing.
Stateflow, closely tied to Simulink, is adequate for logic and state-based testing but shows limited interoperability, making it best as a Simulink complement.
PREEvision is tailored to automotive applications, offering strong traceability, standards compliance, and scalability. While powerful in system architecture modeling and a solid automation framework, its auto-generation is Medium.
Eclipse Papyrus benefits from strong interoperability and model customization via open standards. However, it emphasizes research and modeling over automation and performance.
IBM RSA supports traceability and enterprise integration with moderate automation and scripting. It lacks high-throughput testing, making it suitable for projects prioritizing traceability over automation intensity.
GraphWalker specializes in model-based test generation and execution, with strong CI/CD and scripting support. It has limited scalability and system integration focus on specialized automated testing scenarios, see Table~\ref{tab:cc_modeltools}.

With 15 criteria evaluated at a maximum of 2 points each, the highest attainable score per tool is 30. The aggregated scores (Σ row) yield the following ranking: Simulink (28) > OpenMETA (23) > PREEvision (22) > GraphWalker (20) > Stateflow = IBM RSA (17) > Papyrus (15). Simulink demonstrates the most comprehensive criterion coverage, while OpenMETA and PREEvision emerge as leading tools in multidisciplinary design and automotive-specific modeling, respectively.
\begin{keytakeaway}
\textbf{Key Takeaway.}
With 28/30 points (15 criteria, High=2), Simulink remains the
benchmark for model‑based development, excelling in automation,
scalability, and real‑world simulation, while integrating smoothly into
CI/CD pipelines.
OpenMETA scores 23/30; It offers strong integration,
customization, and automation, making it well-suited to multidisciplinary design analysis and optimization.
PREEvision follows at 22/30, providing robust traceability,
automotive standards compliance, and high scalability, though its automated test‑case generation is only a medium.
GraphWalker achieves 20/30, focusing on model‑based testing
generation, CI/CD, and scripting, but scales only moderately and offers a limited unified environment.
Stateflow and IBM RSA each reach 17/30: Stateflow complements Simulink for state logic yet lags in interoperability, while RSA emphasizes enterprise traceability over high‑throughput execution.
Papyrus ranks last at 15/30, reflecting its research‑oriented
feature set and modest automation support despite strong interoperability.
No single modeling tool covers all 15 criteria at High level; pairing
Simulink with a specialist, such as OpenMETA or GraphWalker, can extend
coverage without redundant functionality.
\end{keytakeaway}

\subsubsubsection{Data Management / Test Platform}
\begingroup
\footnotesize
\setlength\LTpre{0pt}\setlength\LTpost{0pt}

\begin{table}[htbp]
\renewcommand\arraystretch{1.05}
  \centering
  \begin{tabularx}{\textwidth}{@{}
    >{\centering\arraybackslash}p{1.2cm}   
    >{\raggedright\arraybackslash}X          
    *{7}{>{\centering\arraybackslash}p{0.85cm}}  
  @{}}                                       
    \toprule
 \rot{\textbf{Prioritized ID}} & \textbf{Criteria} &
      \rot{\textbf{Doors}} &
      \rot{\textbf{Jenkins}} &
      \rot{\textbf{Cucumber}} &
      \rot{\textbf{FitNesse}} &
      \rot{\textbf{UDE}} &
      \rot{\textbf{code\-Beamer}} &
      \rot{\textbf{CANalyzer}} \\
    \midrule
    \multicolumn{9}{@{}l}{\textbf{Testing Tools and Infrastructure Requirements}}\\[1pt]
    C$_{5}$P$_{1}$ & Interoperability and open standards           & \fullcirc & \fullcirc & \fullcirc & \fullcirc & \halfcirc & \fullcirc & \fullcirc \\
    C$_{6}$P$_{1}$ & Scalability and performance                   & \fullcirc & \fullcirc & \halfcirc & \halfcirc & \fullcirc & \fullcirc & \fullcirc \\
    C$_{8}$P$_{1}$  & End-to-end traceability     & \fullcirc & \halfcirc & \halfcirc & \halfcirc & \fullcirc & \fullcirc & \fullcirc \\
    \addlinespace[1pt]
    \multicolumn{9}{@{}l}{\textbf{Test Automation and Execution}}\\[1pt]
    C$_{17}$P$_{1}$  & Automation framework capabilities                      & \halfcirc & \fullcirc & \fullcirc & \fullcirc & \fullcirc & \fullcirc & \fullcirc \\
    C$_{18}$P$_{1}$ & Automated test case generation               & \emptycirc & \halfcirc & \halfcirc & \halfcirc & \halfcirc & \halfcirc & \halfcirc \\
    C$_{19}$P$_{1}$ & High-throughput and parallel execution        & \emptycirc & \fullcirc & \halfcirc & \halfcirc & \fullcirc  & \halfcirc & \fullcirc \\
    C$_{20}$P$_{1}$& CI/CD integration                            & \emptycirc & \fullcirc & \fullcirc & \fullcirc & \halfcirc  & \fullcirc & \halfcirc \\
    C$_{21}$P$_{1}$ & Robust scripting and error handling           & \halfcirc  & \fullcirc & \halfcirc & \halfcirc & \fullcirc  & \halfcirc & \fullcirc \\
    \addlinespace[1pt]
    \multicolumn{9}{@{}l}{\textbf{Advanced Techniques and Technologies}}\\[1pt]
    C$_{22}$P$_{2}$ & Integration of advanced test techniques       & \halfcirc & \fullcirc & \fullcirc & \fullcirc & \fullcirc & \fullcirc & \fullcirc \\
    C$_{23}$P$_{2}$ & AI-driven optimization                       & \fullcirc & \fullcirc & \halfcirc & \halfcirc & \halfcirc & \fullcirc & \fullcirc \\
    C$_{24}$P$_{1}$ & Simulation environments        & \halfcirc & \fullcirc & \halfcirc & \halfcirc & \fullcirc  & \fullcirc & \fullcirc \\
\midrule[0.3pt]                   
\textbf{$\Sigma$} & \multicolumn{1}{r}{\textbf{Score}} &
\textbf{12} & \textbf{20} & \textbf{15} & \textbf{15} &
\textbf{18} & \textbf{19}& \textbf{20} \\[-0.1em]
\bottomrule

  \end{tabularx}
  \caption{Coverage of prioritized criteria from three categories by seven data management / test platform tools. Ratings are High (\fullcirc), Medium (\halfcirc), Low (\emptycirc), or not applicable (–). The bottom row $\Sigma$ lists the aggregated scores per technique, where High = 2, Medium = 1, Low =0.}

  \label{tab:cc_data_tools}
\end{table}
\endgroup
DOORS excels in traceability, collaboration, and requirements management, with strong support for compliance, metrics, reporting, and feedback loops. Furthermore, DOORS offers AI add-ons such as ReqIF analytics. It lacks robust CI/CD integration, scripting, and automated test case generation, making it best suited for regulated environments prioritizing traceability and oversight.
Jenkins offers strong automation, scalability, and CI/CD support but provides limited requirements management, traceability, or structured collaboration functionality. It is well-suited for Agile/DevOps teams focused on automating build, test, and deployment pipelines. Furthermore, Jenkins also provides plug-ins for ML-based test results. 
Cucumber supports BDD, collaboration, and NL-based automated testing. However, it offers limited metrics, dashboards, and formal role management support. It is ideal for agile teams needing clear communication of requirements and test automation.
FitNesse provides solid automation, scripting, and CI/CD integration support, but lacks scalability for large test suites.
UDE performs well in embedded systems testing and scales to large embedded test benches, offering strong traceability and scripting for automation.
codeBeamer is highly rated for life cycle management with its built-in automation engine, but provides only moderate support for fully automated test generation. CANalyzer is a standard in automotive network analysis, diagnostics, simulation, and embedded test automation, particularly excelling in traceability, see Table~\ref{tab:cc_data_tools}.

Across the 11 assessed criteria (maximum 22 points), Jenkins and CANalyzer share the lead with 20/22, followed by codeBeamer 19, UDE 18, Cucumber and FitNesse 15, and DOORS 12.
The ranking highlights Jenkins’ breadth in automation and CI/CD, whereas CANalyzer scores through deep automotive protocol support and simulation.

\begin{keytakeaway} \textbf{Key Takeaway.} With 20/22 points, Jenkins offers the broadest criterion coverage, pairing full‑stack automation with rich plug‑ins for CI/CD, AI‑assisted result triage, and performance analytics.CANalyzer also scores 20/22, excelling in automotive network diagnostics, simulation environments, and end‑to‑end traceability. codeBeamer follows at 19/22, combining life cycle management, traceability, and a built‑in automation engine, though its auto‑generation of test cases is only medium. 

UDE achieves 18/22 and is well suited to large embedded test‑benches thanks to high scalability, strong scripting, and solid traceability. Cucumber and FitNesse each reach 15/22, providing readable BDD or table‑driven automation but limited analytics and throughput. DOORS ranks last at 12/22, reflecting its focus on requirements traceability over advanced automation. 

No single platform scores High on every criterion; pairing a CI‑centric tool such as Jenkins with a domain specialist like CANalyzer or codeBeamer can yield broader capability without redundant overlap. \end{keytakeaway}
\subsubsubsection{Quality Management Software}
\begingroup
\footnotesize
\setlength\LTpre{0pt}\setlength\LTpost{0pt}

\begin{table}[htbp]
\renewcommand\arraystretch{1.05}
  \centering
  \begin{tabularx}{\textwidth}{@{}
      >{\centering\arraybackslash}p{1.2cm}  
      >{\raggedright\arraybackslash}X           
      *{4}{>{\centering\arraybackslash}p{0.85cm}}  
  @{}}                                          
    \toprule
  \rot{\textbf{Prioritized ID}} & \textbf{Criteria} &
      \rot{\textbf{ALM}} &
      \rot{\textbf{Spira\-Test}} &
      \rot{\textbf{Marama\-AIC}} &
      \rot{\textbf{TestME\-Req}} \\
    \midrule
    \multicolumn{6}{@{}l}{\textbf{Test Planning and Scope Management}}\\[1pt]
   C$_{1}$P$_{1}$ & Model-based approach                       & --        & --        & \fullcirc & \fullcirc \\
    \addlinespace[1pt]

    \multicolumn{6}{@{}l}{\textbf{Testing Tools and Infrastructure Requirements}}\\[1pt]
    C$_{4}$P$_{1}$ & Unified and integrated environment          & \halfcirc & \halfcirc & \emptycirc & \emptycirc \\
    C$_{5}$P$_{1}$  & Interoperability and open standards         & \halfcirc & \halfcirc & \halfcirc  & \halfcirc \\
    C$_{8}$P$_{1}$& End-to-end traceability             & \halfcirc & \halfcirc & \emptycirc & --        \\
    \addlinespace[1pt]

    \multicolumn{6}{@{}l}{\textbf{Requirements and Specification Quality}}\\[1pt]
    C$_{15}$P$_{1}$ & Completeness and Consistency             & --        & --        & \halfcirc & --        \\
    \addlinespace[1pt]

    \multicolumn{6}{@{}l}{\textbf{Reporting, Metrics, and Continuous Improvement}}\\[1pt]
    C$_{26}$P$_{3}$& Reporting dashboards       & \fullcirc & \fullcirc & --        & --        \\
    C$_{27}$P$_{2}$ & Automated analysis and visualization       & \fullcirc & \fullcirc & --        & --        \\
    \addlinespace[1pt]

    \multicolumn{6}{@{}l}{\textbf{Organizational Processes, Collaboration, and Agile Practices}}\\[1pt]
    C$_{29}$P$_{1}$ & Stakeholder alignment and communication         & \emptycirc & \emptycirc & --       & --        \\   
    C$_{31}$P$_{2}$ & Cross-functional collaboration      & \halfcirc & \halfcirc & --        & --        \\
    C$_{32}$P$_{2}$ & Defined roles and responsibilities         & \emptycirc & \emptycirc & --       & --        \\
    C$_{33}$P$_{2}$ & Continuous process improvement & \emptycirc & \emptycirc & --     & --        \\
    C$_{34}$P$_{2}$ & Knowledge sharing          & \emptycirc & \emptycirc & --       & --        \\

\midrule[0.3pt]                   
\textbf{$\Sigma$} & \multicolumn{1}{r}{\textbf{Score}} &
\textbf{8} & \textbf{8} & \textbf{4} & \textbf{3} \\[-0.1em]
\bottomrule

  \end{tabularx}
  \caption{Coverage of prioritized criteria from five categories of the Criteria Catalog by four quality management tools. Ratings are High (\fullcirc), Medium (\halfcirc), Low (\emptycirc), or not applicable (–). The bottom row $\Sigma$ lists the aggregated scores per technique, where High = 2, Medium = 1, Low =0.}
  \label{tab:cc_qualitytools}
\end{table}
\endgroup
ALM and SpiraTest provide extensive reporting and dashboards. MaramaAIC and TestMEReq are academic tools with prototypical implementations. While they show promise, their industrial integration is limited, and they may not yet fully address complex real-world scenarios, see Table~\ref{tab:cc_qualitytools}.

Across 12 assessed criteria (maximum = 24 points), ALM and SpiraTest tie for the lead with 8/24, followed by MaramaAIC 4 and TestME-Req 3. The low absolute numbers show that commercial QMS suites deliver strong reporting and dashboards but cover only a subset of the broader criteria catalogue. At the same time, the two academic tools remain prototypes with limited industrial alignment.

\begin{keytakeaway} \textbf{Key Takeaway.} ALM and SpiraTest share the top score (8/24, 12 criteria, High = 2), driven by comprehensive dashboards (C$_26$) and automated visualisation (C$_27$). MaramaAIC scores 4/24: its model-based approach shows promise for research settings but lacks a unified environment and full traceability. TestME-Req ranks last at 3/24, reflecting its early-stage focus on requirements modeling with minimal infrastructure support. 
Publicly available documentation for these tools is limited; several tool–criterion combinations could not be rated and are marked as not applicable (–). Overall, quality management tools excel in reporting but fall short on automation, collaboration, and continuous-improvement criteria; pairing a QMS suite such as ALM or SpiraTest with a CI platform or life cycle tool can close the gap for enterprise projects. \end{keytakeaway}

\section{Threats to Validity}
\label{threats}
This review focuses on peer-reviewed literature retrieved from major digital libraries and is therefore not a full multi-vocal literature review. It does not systematically include grey literature such as internal technical reports, white papers, or non-archival experience reports. 
As a result, specific industrial practices, especially those documented only in internal or non-indexed sources, may be underrepresented.
Consequently, our synthesis is grounded in published academic and joint academic–industry studies and may lag behind rapidly evolving industrial practice, especially in organizations that publish little or not at all. The criteria catalog should therefore be interpreted as a structured synthesis of published work complemented by practitioner experience, rather than as a statistically representative survey of the global automotive industry.
Restricting the search to English-language sources may have excluded relevant studies from countries with strong automotive sectors (e.g., Japan, China, Korea, and others), which can limit the geographical diversity of the evidence base.

Many primary studies rely on case studies or surveys with modest sample sizes, often within a single organization or a limited ecosystem. As a result, our synthesized findings should be interpreted as analytic generalizations rather than statistically representative measures of the global automotive industry. Where conclusions are grounded in only a small number of PSs, we explicitly note this in the text so that readers can judge the strength and scope of the underlying evidence.

The criteria catalog is expert-derived and reflects the judgment of the authors and consulted practitioners. Although this is common for early criteria catalogs, it introduces subjectivity. We have not yet conducted a structured validation study with a broader pool of practitioners, for example, through surveys or workshops. Such validation is left for future work.

In addition, the review is subject to several standard SLR biases. Retrieval bias may arise because some relevant work is not indexed in the selected databases or does not match our controlled vocabulary. Publication bias is likely, since studies with negative or inconclusive results are less frequently reported in peer-reviewed venues. Selection and data extraction bias can occur because screening and coding were performed manually, despite the use of explicit inclusion and exclusion criteria, pilot calibration, and consensus-based resolution of borderline cases.
\section{Synthesis and Answers to the Research Questions}
\label{sec:rq-answers}

\noindent\textbf{RQ1. Which testing-related challenges are reported for automotive software and cyber-physical systems across the full test life cycle?} 
Our review consolidates the reported challenges into nine recurring areas across the life cycle: 
(i) weak and late requirements combined with poor traceability and reliance on natural language and testers expertise (ambiguity, inconsistency, incomplete specifications); 
(ii) variability and configuration explosion; 
(iii) fragmented toolchains and poor interoperability; 
(iv) limited, brittle, or late CI/CD integration and resource-constrained XiL; 
(v) NL test specifications that hinder reuse and automation; 
(vi) system-of-systems and IoT scale with flaky connectivity and data issues; 
(vii) limited adoption of advanced test design and prioritization at scale; 
(viii) semantic gaps across MBSE artifacts and timing semantics; 
(ix) organizational factors (roles, planning, reporting, knowledge sharing). 
Evidence for these areas appears in Section~\ref{sec:soa}.

\medskip
\noindent\textbf{RQ2. What technical and process requirements have been proposed to improve efficiency and effectiveness?} 
We synthesize the improvement directions into seven requirement categories from our Criteria Catalog (Table~\ref{tab:c_table}), with P$_1$ priorities in parentheses:
(1) Test planning and scope management (model-based approach; variants and configuration management; requirements alignment) (P$_1$);
(2) Testing tools and infrastructure (unified environments; open standards and APIs; scalability; traceability) (P$_1$);
(3) Requirements quality (unambiguous, verifiable, complete; better NL tooling and formalization paths) (P$_1$);
(4) Test automation (maintainable DSLs; stable CI agents; virtualization and HiL orchestration) (P$_1$);
(5) Advanced techniques and technologies (AI-assisted prioritization and coverage, formal methods where feasible, realistic simulation) (P$_2$–P$_3$);
(6) Reporting and metrics (APFED, FDR, CE, ROI dashboards; automated analytics) (P$_2$);
(7) Organizational processes and agile practices (clear roles, cross-functional collaboration, knowledge sharing, continuous improvement) (P$_2$). 
These categories and priorities are derived from the 34 criteria consolidated in Section~\ref{sec:requirements}. 

In particular, the category of advanced techniques and technologies aggregates approaches such as AI-assisted testing, formal methods, and high-fidelity simulation that are frequently discussed in the literature but, according to both the primary studies and our insights from industrial collaborations, still exhibit limited and uneven industrial adoption; we therefore interpret them as promising but emerging directions rather than established mainstream practice.

In practice, practitioners are expected to adapt and prioritize the catalog according to their local context, constraints, and responsibilities, rather than applying all criteria uniformly.

\medskip
\noindent\textbf{Challenge-to-requirement mapping.} Table~\ref{tab:rq-crosswalk} links the RQ1 challenges to the RQ2 requirement directions to make the practical implications explicit.

\begingroup
\small
\rowcolors{2}{white}{gray!5}
\begin{table}[htbp]
\caption{RQ1 challenges mapped to RQ2 requirement directions}
\label{tab:rq-crosswalk}
\centering
\begin{tabularx}{\textwidth}{>{\raggedright\arraybackslash}p{4.5cm} >{\raggedright\arraybackslash}X}
\toprule
\textbf{Challenge areas} & \textbf{Requirement directions} \\
\midrule
Ambiguous, late, or weak requirements & Requirements quality uplift; NL assistance; selective formalization; traceability to tests (C$_3$,C$_{14}$--C$_{18}$). \\
Variability and configuration explosion & Model-based planning; variant management and test selection (C$_1$–C$_2$). \\
Tool fragmentation and poor interoperability & Unified environment; open standards and APIs; end-to-end traceability (C$_4$–C$_7$). \\
Brittle CI/CD and costly XiL & Virtualization; resource scheduling; pipeline hardening and metrics (C$_8$–C$_{13}$, C$_{26}$--C$_{28}$). \\
Natural-language test specs hinder reuse & Test DSLs; pattern libraries; template conformance (C$_{10}$C$_{12}$). \\
System-of-systems and IoT complexity & Robust simulation, fault isolation, data quality, and security in test environments (C$_{24}$). \\
Limited adoption of advanced test design & AI-assisted prioritization and coverage; tester-in-the-loop MBT (C$_{23}$). \\
Semantic gaps across MBSE artifacts & Standardized modeling, composition semantics, and timing treatments (C$_1$, C$_5$). \\
Organizational and process issues & Clear roles; cross-functional collaboration; knowledge sharing; continuous improvement (C$_{29}$--C$_{34}$). \\
\bottomrule
\end{tabularx}
\end{table}
\endgroup

Summarizing, RQ1 reveals a consistent pattern of requirements quality and traceability gaps, tool and process fragmentation, and scaling limits in CI and XiL that together impede efficiency and reliability. Several of these challenge families, such as requirements quality and traceability or tool and process fragmentation, cut across multiple SWEBOK knowledge areas (requirements, design, testing, and maintenance), which underlines their systemic and interdisciplinary nature. RQ2 is addressed by a prioritized criteria catalog that emphasizes model-based planning, open and interoperable toolchains with end-to-end traceability, uplift of requirements quality and NL support, pragmatic automation and virtualization, targeted use of AI and formal methods, actionable metrics, and lightweight organizational practices that enable continuous improvement.

The resulting mapping of specification techniques and tools to the criteria catalog is descriptive and analytical; it supports structured selection and gap analysis but does not provide a quantitative benchmark of comparative effectiveness among tools or techniques.
The criteria catalog and challenge–requirement mappings synthesize evidence from 50 peer-reviewed primary studies, complemented by practitioner experience and observations from several industrial collaborations.
The challenge patterns identified for RQ1 recur across several PSs, although often in case-study settings with limited sample sizes.
\section{Concluding Remarks}
\label{sec:conclusion}
Modern vehicles are intricate systems composed of tightly integrated hardware and software developed by a wide range of organizations. This complexity makes it nearly impossible to predict system behavior in every scenario, significantly increasing the challenges of testing and validation. Yet, thorough testing is crucial to meet stringent safety regulations and legal standards, as well as to satisfy user expectations for reliability, performance, and seamless functionality. As the industry evolves, robust and adaptive testing strategies will remain vital to ensuring the quality and trustworthiness of automotive products.

In line with these needs, we identified testing process challenges in the automotive industry by conducting a comprehensive SLR supported by the snowballing method. We reviewed 50 primary studies and extracted crucial criteria for an improved testing methodology. Using the PRISMA framework, we collected test case specification techniques and various testing tools and evaluated their suitability based on a defined criteria catalog.
  
Our results highlight a critical need for testing strategies that integrate seamlessly across the automotive development life cycle. Addressing the identified challenges requires embracing standardized, model-based approaches, enhancing automation capabilities, and ensuring tool interoperability. Such advancements can significantly increase testing efficiency, reduce development costs, and accelerate time-to-market for automotive innovations. Moreover, closely aligning testing processes with evolving software requirements and leveraging CI practices can enhance safety standards and facilitate regulatory compliance.

Our analysis indicates that while automotive software testing employs established industry standards and tools, it still heavily relies on manual intervention and individual expertise, suggesting that the field is transitioning from an early stage of maturity to a more established industrial practice. A gap remains between the theoretical frameworks proposed by academic research and their practical application. Bridging this gap through collaborative frameworks, technology transfers, and improved feedback loops is crucial for advancing both theoretical knowledge and industrial applicability. 

The present research contributes to the field by systematizing current approaches, standards, and tools, and by identifying critical areas for improvement across the test life cycle. By closing academia–industry gaps through targeted collaboration and adopting standardized, automated, and integrated testing practices, the automotive sector can gain a substantial competitive edge in an increasingly software-defined market.  

In practice, we expect organizations to adopt the proposed criteria catalog incrementally and to integrate it with heterogeneous, legacy-constrained toolchains, rather than replacing them with a single unified methodology.

An important direction for future work is empirical benchmarking of selected specification techniques and toolchains against the proposed criteria catalog to evaluate their effectiveness and scalability under realistic industrial conditions.
A dedicated multi-vocal literature review that systematically incorporates grey literature and industry-internal documents would be a valuable next step to strengthen further and broaden the evidence base for these findings.
Emerging domains, such as V2X communication, connected IoT services, and higher levels of driving automation, introduce additional testing challenges, which we highlight at a high level here. A more detailed treatment of these domains is an important direction for subsequent reviews.

\appendix
\section{Automotive Protocols and Standards}
\label{sec:background}
ASPICE (Automotive Software Process Improvement and Capability dEtermination) is a framework to evaluate and improve the software development process, specifically in the automotive industry. It includes a Process Assessment Model (PAM) intended for conformant assessments of process capability in developing embedded automotive systems, developed by ISO/IEC 33004:2015. ASPICE has its own Process Reference Model (PRM), which is based on the ASPICE Process Reference Model 4.5 and is further tailored to meet the specific needs of the automotive industry. For processes beyond the scope of ASPICE, it allows integrating processes from other models, such as ISO/IEC 12207 or ISO/IEC/IEEE 15288, depending on organizational needs.

\begin{figure*}
	\centering
	\includegraphics[width=1\linewidth]{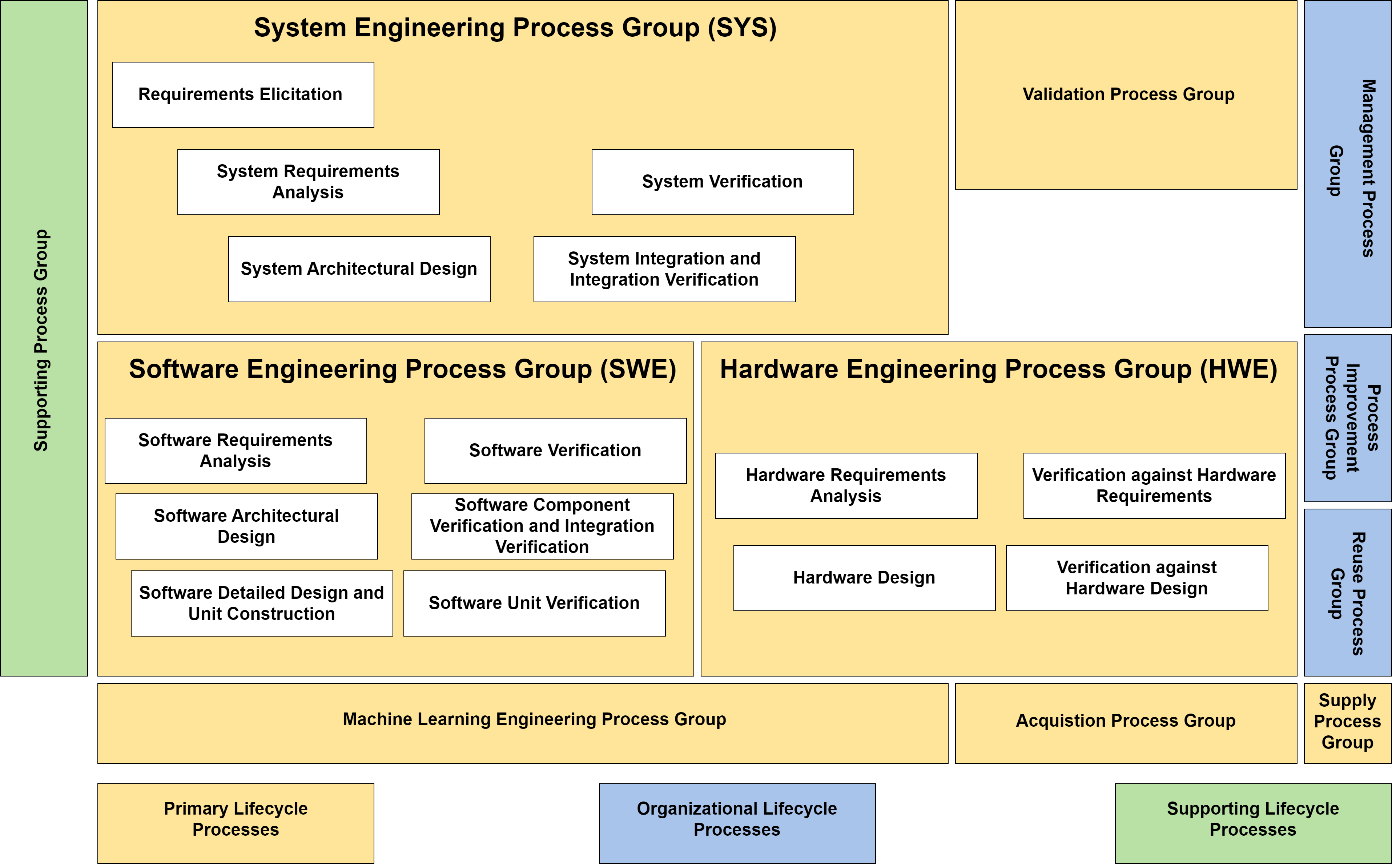}
	\caption{ASPICE Process Reference Model (PRM)}
	\label{fig:aspice}
\end{figure*}

The PRM is organized into three categories: Primary Life Cycle Processes, Organizational Life Cycle Processes, and Supporting Life Cycle Processes, as illustrated in Figure~\ref{fig:aspice}.
The Primary Life Cycle Processes consist of processes that may apply to an acquirer of products from a supplier or product development, including the engineering processes needed for specification, design, implementation, integration, and verification. This paper focuses on the Primary Life Cycle Processes. The System Engineering Process Group (SYS) encompasses a suite of processes focused on the elicitation and management of both customer and internal requirements, as well as the formulation of system architecture and the execution of integration and verification at the system level. Following the SYS, the Validation Process Group (VAL) is created to provide evidence that the product met its intended usage expectations. 

This initial phase is crucial for establishing the framework for developing and applying specific testing protocols. It systematically addresses eliciting and managing requirements from customers and internal stakeholders. It sets the groundwork for defining system architecture and integration and verification at the system level. Following the establishment of this foundational framework, the Software Engineering Process Group (SWE) and Hardware Engineering Process Group (HWE) further refine the engineering and testing processes, focusing on the specific needs of software and hardware development, respectively~\cite{aspice}.

AUTOSAR (Automotive Open System Architecture) is a standard for software and methodology enabling open E/E system architecture for the automotive industry. AUTOSAR aims to fulfill future vehicle requirements and simplify software development and vehicle integration by providing a standard platform enabling seamless communication between different ECUs. AUTOSAR distinguishes between the Classic Platform and the Adaptive Platform. 
The AUTOSAR Classic Platform focuses on standardization and compatibility across different vehicle domains. It runs three software layers on a microcontroller: application, runtime environment (RTE), and basic software (BSW)~\cite{autosar}. 
The AUTOSAR Adaptive Platform is designed to support the development of highly complex and dynamic applications~\cite{Stawski}. It implements the AUTOSAR Runtime for Adaptive Applications (ARA). Services and APIs are available interfaces. The platform consists of functional clusters. The clusters are grouped into services and the Adaptive AUTOSAR Basis~\cite{autosar}. 
Furthermore, the standardized application interface is crucial in both platforms, but the approach differs. While the Classic Platform uses a more rigid set of predefined interfaces for fixed-function ECUs, the Adaptive platform leverages flexible APIs that enable communication between various software modules within or across multiple ECUs in distributed systems~\cite{Stawski}.

\section*{Acknowledgements}
This work was partially funded by the project SofDCar (19S21002, German Federal Ministry for Economic Affairs and Climate Action) and by the EU project HAL4SDV (900027790, European funding initiative Chips JU).

\bibliographystyle{elsarticle-num} 
\bibliography{database}




\end{document}